\documentclass[12pt]{iopart}
\usepackage{hyperref}
\usepackage{graphicx}
\usepackage{xcolor}
\begin{document}
\title[Vertical Growth of van der Waals stacked 2D materials]{Modeling the vertical growth of van der Waals stacked 2D materials using the diffuse domain method}
\author{Zhenlin Guo$^1$,  Christopher Price$^2$, Vivek B. Shenoy$^{2,3,4}$ and John Lowengrub$^{1,5}$}
	\address{$^1$ Department of Mathematics, University of California, Irvine CA 92697-3875, USA}
	\address{$^2$ Department of Materials Science and Engineering, University of Pennsylvania, Philadelphia PA 19104-6272, USA}
	\address{$^3$ Department of Mechanical Engineering \& Applied Mechanics, University of Pennsylvania, Philadelphia PA 19104-6272, USA}
	\address{$^4$ Bioengineering, University of Pennsylvania, Philadelphia PA 19104-6272, USA}
	\address{$^5$ Department of Biomedical Engineering, University of California, Irvine CA 92697-3875, USA}
	\ead{lowengrb@math.uci.edu}


\begin{abstract}
Vertically-stacked monolayers of graphene and other atomically-thin 2D materials have attracted considerable research interest because of their potential in fabricating materials with specifically-designed properties. Chemical vapor deposition has proved to be an efficient and scalable fabrication method. However, a lack of mechanistic understanding has hampered efforts to control the fabrication process beyond empirical trial-and-error approaches. In this paper, we develop a general multiscale Burton-Cabrera-Frank (BCF) type model of the vertical growth of 2D materials to predict the necessary growth conditions for vertical versus in-plane (monolayer) growth of arbitrarily-shaped layers. This extends previous work where we developed such a model assuming the layers were fully-faceted (Ye et al.,  ACS Nano, {\bf 11}, 12780-12788, 2017). To solve the model numerically, we reformulate the system using the phase-field/diffuse domain method that enables the equations to be solved in a fixed regular domain. We use a second-order accurate, adaptive finite-difference/nonlinear multigrid algorithm to discretize and solve the discrete system. We investigate the effect of parameters, including the van der Waals interaction energies between the layers, the kinetic attachment rates, the edge-energies and the deposition flux, on layer growth and morphologies. While the conditions that favor vertical growth generally follow an analytic thermodynamic criterion we derived for circular layers, the layer boundaries may develop significant curvature during growth, consistent with experimental observations. Our approach provides a mechanistic framework for controlling and optimizing the growth multilayered 2D materials.

\end{abstract}


%

\section{Introduction}

Two-dimensional (2D) materials including graphene and transition metal dichalcogenides (TMDs) have garnered unprecedented interest in pursuit of unique electronic, optical, mechanical, and thermal properties \cite{Choi2017,Duong2017,Hong2017,Shi2018,Gobbi2018}. Compared to homogeneous monolayers, multilayered heterostructures contain many more degrees of freedom and thus can be ideal platforms for electronic structure engineering of atomically thin 2D semiconducting materials for novel applications. A key challenge in the realization of vertically integrated 2D layers is their synthesis \cite{Solis-Fernandez2017,Choi2017,Frisenda2018}. Chemical vapor deposition (CVD) has proved to be an efficient and scalable method to grow monolayer 2D materials on a variety of metal substrates \cite{Lu2013,Xia2015,Gong2014,Gong2015}. CVD, however, is a complex process that contains many parameters that influence growth. For example, the growth temperature and the deposition flux have been found to be critical parameters for switching from in-plane (monolayer) to vertically-stacked multilayer growth. In $WS_2$/$MoS_2$ heterostructures on $SiO_2$/$Si$ substrates, high temperatures favor the growth of vertically-stacked multilayers while low temperatures favor monolayer growth \cite{Gong2014}. In graphene, a lower deposition flux (e.g., higher concentrations of $H_2$ in the gas) also tends to favor multilayer growth \cite{Zhang2014,Chen2015}. Determining proper growth parameters is clearly a multivariable problem that until recently was tackled using empirical trial-and-error approaches.

In recent work, we developed a multiscale model of the growth of vertically-stacked 2D materials on a substrate using CVD \cite{Ye2017}. The model, which is of Burton-Cabrera-Frank (BCF) type \cite{BCF1951}, accounts for attachment and diffusion of adatoms, van der Waals (vdW) interactions between the layers and the substrate, and edge energies of the layers. To simplify the system, the layers were assumed to be fully-faceted and so their shapes were constrained to be equilateral polygons (e.g., triangles and hexagons). This work predicted the thermodynamic requirements for growth of vertically-stacked faceted layers. The vdW-BCF model predictions on monolayer vs. multilayer morphologies were validated by comparison with a variety of CVD-synthesized $MX_2$ (M = Mo, W; X = S, Se, Te) single-species samples grown under conditions of varying temperature and precursor flux. 

However, as seen in the experiments in \cite{Ye2017,Lu2013} and in other references, the layers need not to be faceted and can develop significant, and even negative, curvatures. Because the layer morphologies influence growth and the material properties, it is important to accurately predict the layer shapes as well. In this paper, we extend the vdW-BCF model in \cite{Ye2017} to account for arbitrary layer shapes. The resulting system is a highly nonlinear free boundary problem. We analyze the model and derive an analytic thermodynamic criterion for vertical growth assuming the layers are circular. To simulate the model when the layer geometries are unconstrained, we develop a second-order accurate phase-field/diffusion-domain method (DDM) that enables us to simulate the system by solving a reformulated system (vdW-BCF-DDM equations) in a fixed regular domain. 

The diffuse-domain, or smoothed boundary, method is an attractive approach for solving partial differential equations in complex geometries because of its simplicity and flexibility. In this method the complex geometry is embedded into a larger, regular domain. The original PDE is reformulated using a smoothed characteristic function of the complex domain and source terms are introduced to approximate the boundary conditions. An advantage of this approach is that the reformulated equations can be solved by standard numerical techniques without requiring body-fitted meshes, additional interfacial meshes or special stencils and the same solver can be used for any geometry. The diffuse-domain method (DDM) was introduced in \cite{Kockelkoren-2003} to solve diffusion equations with Neumann (no-flux) boundary conditions, to PDEs with Robin and Dirichlet boundary conditions in \cite{Diffuse-Domain} and to cases in which bulk and surface equations are coupled \cite{tei09}. Later, in \cite{Yu2012} and \cite{Poulson-2018} alternate derivations of diffuse-domain methods for such problems were presented. In \cite{000356209300006} a matched asymptotic analysis for general DDMs with Neumann and Robin boundary conditions showed that for certain choices of the source terms, the DDMs were second-order accurate in $\epsilon$ and in the grid size $h$ {in both the \(L^2\) and the \(L^\infty\) norms}, taking $\epsilon\propto h$, see the recent paper \cite{Burger-2017} for a rigorous proof. 

In \cite{Ratz2015}, a DDM was proposed to solve a BCF model of epitaxial growth of thin, crystalline films that combined a DDM reformulation of the adatom diffusion equations together with a Cahn-Hilliard-type equation to model the dynamics of the films. This approach considered only isotropic edge energies and kinetic coefficients and did not consider van der Waals interactions. Further, the DDM used in \cite{Ratz2015} did not use a second-order accurate formulation and thus was only first order accurate in $\epsilon$ (and $h$ assuming $\epsilon \propto h$). 

Here, we combine and extend the approaches from \cite{000356209300006} and \cite{Ratz2015} to develop a second-order accurate adaptive finite-difference/nonlinear multigrid method to discretize and solve the vdW-BCF-DDM equations numerically. We investigate the effect of parameters, including vdW interaction energies between the layers, kinetic attachment rates, edge-energies and deposition flux, on layer growth and morphologies. While the conditions that favor vertical growth generally follow the thermodynamic criterion we derived for circular layers, the layer boundaries may develop significant curvature during growth, consistent with experimental observations, that can also influence the growth kinetics.

The outline of the paper is as follows. In Sec. \ref{model}, we present and analyze the vdW-BCF model for arbitrary layer shapes. In Sec. \ref{DDM} we present the phase-field/DDM reformulation of the vdW-BCF model and briefly describe the numerical methods used. In Sec. \ref{NumRes}, we present numerical simulation studies and in Sec. \ref{Conclusion} we present conclusions and discuss future work. Additional details are provided in the Appendices.

\section{The vdW-BCF model for the growth of vertically-stacked multilayers}\label{model}
\begin{figure}
                 \centering
                \includegraphics[width=0.8\textwidth]{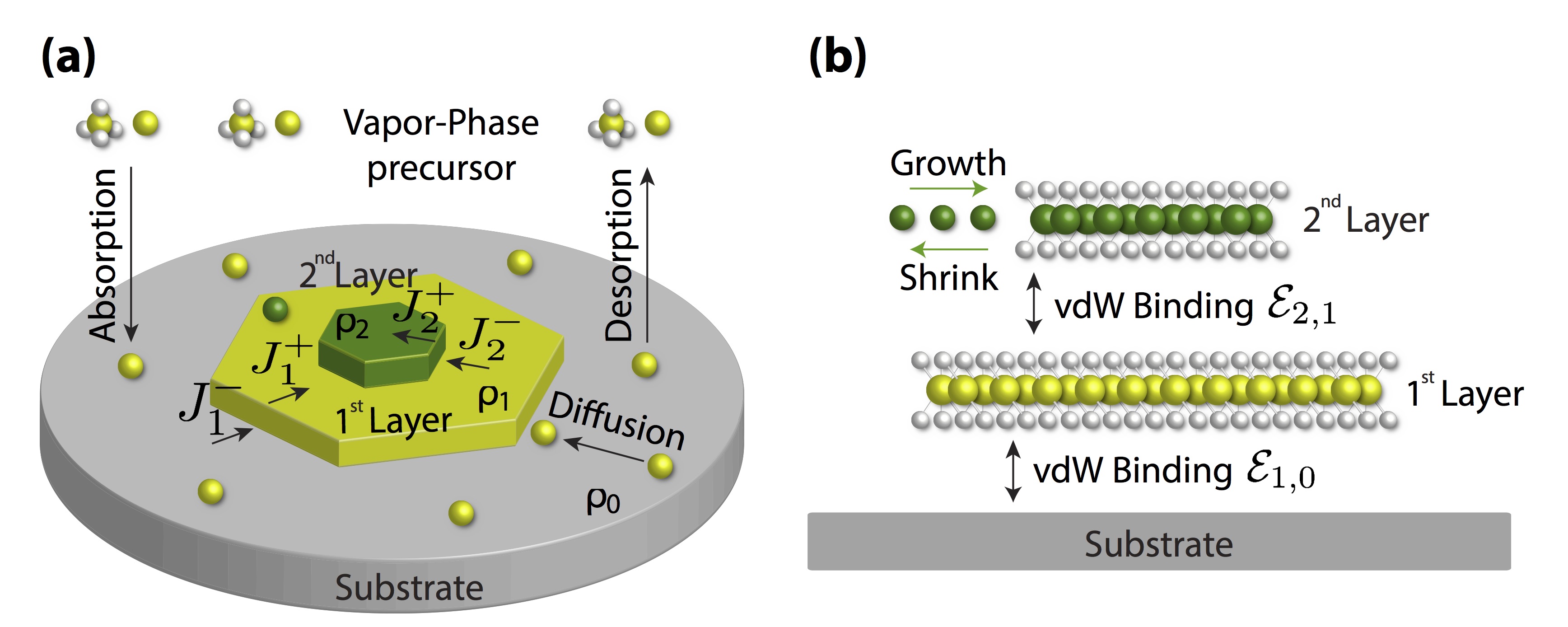}
                       \caption{(a), Schematic of epitaxial growth of 2D materials; (b) Schematic of vdW interactions between the layers and the substrate.}
\label{diagram figure}
\end{figure}

Let $\Omega_0$ denote the substrate, $\Omega_1$ denote a layer of atomic height 1 and $\Omega_2$ be a layer of atomic height 2 with boundaries $\Gamma_0$, $\Gamma_1$ and $\Gamma_2$, respectively. See the diagrams in Figs. \ref{diagram figure} and \ref{diffuse domain method} (left column). The system free energy is taken to be:
\begin{eqnarray}
 E&=\sum_{i=1}^2\left(-\int_{\Omega_{i}}\mathcal{E}_{i} {\rm d}A +\int_{\Gamma_{i}}\gamma_{i} {\rmd}S\right) \nonumber\\
 &+ k_{B}T\sum_{i=0}^2 \int_{\Omega_{i}} \frac{\rho_{i}}{\rho_{ref}}\ln\frac{\rho_{i}}{\rho_{ref}}+\left(1-\frac{\rho_{i}}{\rho_{ref}}\right)\ln\left(1-\frac{\rho_{i}}{\rho_{ref}}\right){\rmd}A,
\label{free energy}
\end{eqnarray}
where $\mathcal{E}_i$ is the binding energy of layer $i$ that accounts for in-plane bonding and any corresponding vdW interactions. In addition, $\gamma_i=\gamma_i(\theta_i)$ is the edge energy of layer $i$, $\theta_i$ is the normal angle of layer $i$ (e.g., the angle between the normal vector $\mathbf{n}_{\Gamma_i}$, which points into $\Omega_{i-1}$, and the $x$-axis). The function $\rho_i$ is the adatom concentration on layer $i$ and $\rho_{ref}=\Omega_s$ is the concentration of atomic sites (assumed to be the same on the layers). Further, $k_B$ is Boltzmann's constant, $T$ is the temperature and the third term in Eq. (\ref{free energy}) represents the regular solution model free energy.

\subsection{Model equations}
\label{model equations}
By requiring mass to be conserved and that the free energy is non-increasing in time, we can derive a thermodynamically-consistent Burton-Cabrera-Frank (BCF)-like system of equations that govern the dynamics of the adatom densities and the layer morphologies and sizes. Here, we only present the nondimensional equations that include several simplifications. A detailed derivation of the equations, a description and justification of the simplifications and the nondimensionalization are given in \ref{model-derivation}.

The nondimensional adatom concentrations satisfy the diffusion equations
\begin{equation}
\partial_{t} \rho_{i}= D_{i}\Delta \rho_{i} +F_{i} -\tau^{-1}_{d,i} \rho_{i}   ~~~~{\rm in}~~\Omega_{i}, ~~~~i=0,1,2,\label{bcf-double-layer}
\end{equation}
where $D_{i}>0$ is a dimensionless diffusion coefficient, $F_{i}$ is a dimensionless deposition flux and $\tau^{-1}_{d,i}$ a dimensionless desorption rate. These are all assumed to be constant. At the layer boundaries $\Gamma_{2}$ and $\Gamma_{1}$ mass conservation is imposed, which yields the kinetic boundary conditions:
\begin{eqnarray}
q_2^+&=-D_2{\bf{\nabla}}\rho_2\cdot\mathbf{n}_{\Gamma_2}-\rho_2|_{\Gamma_2}v_2= k_{2}^{+}\bigg({\rho}_{2} - \rho^*(-\mathcal{E}_{2}+\mathcal{E}_{1}+\tilde\gamma_{2}\kappa_2)\bigg),\label{BCF-bc11} \\
q_{2}^{-}&=D_1{\bf{\nabla}}\rho_{1}\cdot {\bf{n}}_{\Gamma_{2}}+\rho_1|_{\Gamma_2}v_2 = k_{2}^{-}\bigg(\rho_{1}-\rho^*(-\mathcal{E}_{2}+\mathcal{E}_{1}+\tilde\gamma_{2}\kappa_{2})\bigg),\label{BCF-bc21} \\
q_{1}^{+}&:=-D_1{\bf{\nabla}}\rho_{1}\cdot {\bf{n}}_{\Gamma_{1}}-\rho_1|_{\Gamma_1}v_1  =k_{1}^{+}\bigg({\rho}_{1}- \rho^*(-\mathcal{E}_{1}+\tilde\gamma_{1}\kappa_{1})\bigg),\label{BCF-bc31} \\
q_{1}^{-}&:=D_0{\bf{\nabla}}\rho_{0}\cdot {\bf{n}}_{\Gamma_{1}}+\rho_0|_{\Gamma_1}v_1= k_{1}^{-}\bigg(\rho_{0}-\rho^*(-\mathcal{E}_{1}+\tilde\gamma_{1}\kappa_{1})\bigg).\label{BCF-bc41} 
\end{eqnarray}
Here, $q_{i}^{\pm}$ are the diffusion fluxes of adatoms to the layer boundaries, with the $"+"$ and $"-"$ subscripts denoting limits from the $i^{th}$ and $i-1$ layers, $\rho^{*}$ is a nondimensional measure of the thermodynamic equilibrium density, $\tilde\gamma_{i}=\gamma_i(\theta_i)+\gamma_i''(\theta_i)$, where the primes denote derivatives with respect to $\theta_i$, denotes the layer boundary (edge) stiffness, and $\kappa_{i}$ is the curvature of the edge $\Gamma_{i}$ ($i=1,2$). The constants $k_{i}^{\pm}$ are the dimensionless rates for attachment of adatoms to the edges from the $i$th ($k_i^+$) and $i-1$ ($k_i^-$) layers, respectively. The normal velocity of each layer boundary $\Gamma_{i}$ is given by
\begin{eqnarray}
v_{i}=q_{i}^{+}+q_{i}^{-}+\beta \partial_{s}^{2}\kappa_{i},\label{normal-vel-multi}
\end{eqnarray}
where the dimensionless constant $\beta$ is related to the mobility of an adatom along a curved edge. At the boundary of the substrate, we assume there is no flux of adatoms: $\nabla\rho_0\cdot\mathbf{n}_{\Gamma_0}=0$.

\subsection{Analysis of vdW-BCF model: Radial solutions and growth criteria}\label{model-analysis}
For simplicity, we consider a configuration in which the two layers and substrate are circular and centered at the same point $O$. We assume that the edge energy and the kinetic coefficients are isotropic. We solve the system (\ref{bcf-double-layer})-(\ref{normal-vel-multi}) analytically to derive necessary and sufficient conditions for the growth of layer 2. The layers $\Omega_1$ and $\Omega_2$ have radii $R_1(t)$ and $R_2(t)$. The substrate has radius $R_\infty$, which is fixed. We assume that initially $0 < R_2(0)<R_{1}(0)< R_{0}$ and that the dynamics are dominated by diffusion so that the time derivative on the left hand side of Eq.  (\ref{bcf-double-layer}) is set to zero (quasi-steady case). We further assume the desorption of adatoms is small and so we set $\tau^{-1}_{d,i}=0$. The reduced system can be solved analytically. Here, we present only the results, a complete derivation of the solutions is provided in \ref{model-analysis details}.



The analytical solutions for the densities $\rho_i$ are:
\begin{eqnarray}
{\rho} _{2}&= - \frac{F_{2}}{4D_{2}}r^{2} + A_{2}{\rm ln}(r) +B_{2},\hspace{5mm}0<r<R_{2},\nonumber\\
{\rho} _{1}&= - \frac{F_{1}}{4D_{1}}r^{2} + A_{1}{\rm ln}(r) +B_{1},\hspace{5mm}R_{2}<r<R_{1},\nonumber\\
{\rho} _{0}&= - \frac{F_{0}}{4D_{0}}r^{2} + A_{0}{\rm ln}(r) +B_{0},\hspace{5mm}R_{1}<r<R_\infty,\label{analytical-solution-re}
\end{eqnarray}
where $A_{i}$ and $B_{i}$ are given in \ref{model-analysis details}. When the flux of adatoms is only non-zero on the substrate (e.g., $F_0\ge 0$, $F_2=F_1=0$), which reflects the catalytic decomposition of $CH_4$ vapor on the substrate surface into mobile radicals (e.g., $CH$ and $C$) that can attach to the graphene layers \cite{Meca2017}, the normal velocities of the layer boundaries are given by
%
%
%
%
\begin{eqnarray}
v_{1}&=\frac{d}{dt}R_1(t)=\frac{F_0}{2R_1}\left(R_\infty^2-R_1^2\right)-\frac{R_2}{R_1}v_2,
\label{first-layer-velocity-simplified}\\
v_{2}&=\frac{d}{dt}R_2(t)=\frac{D_{1}\rho^*(-\mathcal{E}_{2}+2\mathcal{E}_{1}+\frac{\gamma_{2}}{R_2}-\frac{\gamma_{1}}{R_1})}{R_{2}\bigg(\ln{\frac{R_2}{R_1}}-\frac{D_{1}}{k^{-}_{2}R_2}- \frac{D_{1}}{k^{+}_{1}R_{1}}\bigg)},
\label{second-layer-velocity-simplified}
\end{eqnarray}
where $\gamma_1$ and $\gamma_2$ are isotropic edge energies. The velocities for the more general case with $F_1$ and $F_2$ not necessarily equal to zero can be found in \ref{model-analysis details}.
Define $\mathcal{E}_{2,1}=\mathcal{E}_2-\mathcal{E}_1$ to be the binding energy density between the two layers and $\mathcal{E}_{1,0}=\mathcal{E}_1$ to be the binding energy density between layer 1 and the substrate. The difference between these two energies, 
\begin{equation}
\Delta\mathcal{E}=\mathcal{E}_{2,1}-\mathcal{E}_{1,0}=\mathcal{E}_2-2\mathcal{E}_1,
\label{binding energy difference}
\end{equation}
is the gain in energy by adding atoms to layer 2 instead of layer 1. An analysis of $v_2$ in Eq. (\ref{second-layer-velocity-simplified}) reveals a sufficient condition for the growth of layer 2:
\begin{eqnarray}
\Delta\mathcal{E}>\frac{\gamma_{2}}{R_2}-\frac{\gamma_{1}}{R_1},
\label{criterion}
\end{eqnarray}
since the denominator in Eq. (\ref{second-layer-velocity-simplified}) is always non-positive. This is analogous to the growth criterion derived in \cite{Ye2017} for faceted layers. This condition states that the difference between the binding energies, $\Delta\mathcal{E}$, must be large enough to overcome the energy penalty of increasing the layer perimeter. It follows that if $R_2>R_{2,c}:=\gamma_2/\Delta\mathcal{E}$, then layer 2 always grows, regardless of the size of layer 1. This is analogous to a critical nucleation size. Further, if $\displaystyle{R_1>R_{1,c}:=\frac{\gamma_1}{\gamma_2/R_2-\Delta\mathcal{E}}}$ then layer 2 always shrinks. When $R_2$ is close to $R_{2,c}$, layer 2 may grow due to kinetic effects. That is, $R_2$ may surpass $R_{2,c}$ before $R_1$ surpasses $R_{1,c}$. Whether this occurs depends on the values of the parameters. For example, slowing down the growth of the first layer (e.g., by decreasing $F_0$) or increasing the rate of growth of the second layer (e.g., by increasing $D_1$, $k_{2}^-$ or $k_1^+$) increases the region of kinetically-driven growth. We call $R_{2,k}$ the kinetic critical radius--- that is, if $R_{2,c}>R_2(0)>R_{2,k}$, then the second layer grows due to the kinetics of the system.

\begin{figure}
                 \centering
                \includegraphics[width=0.8\textwidth]{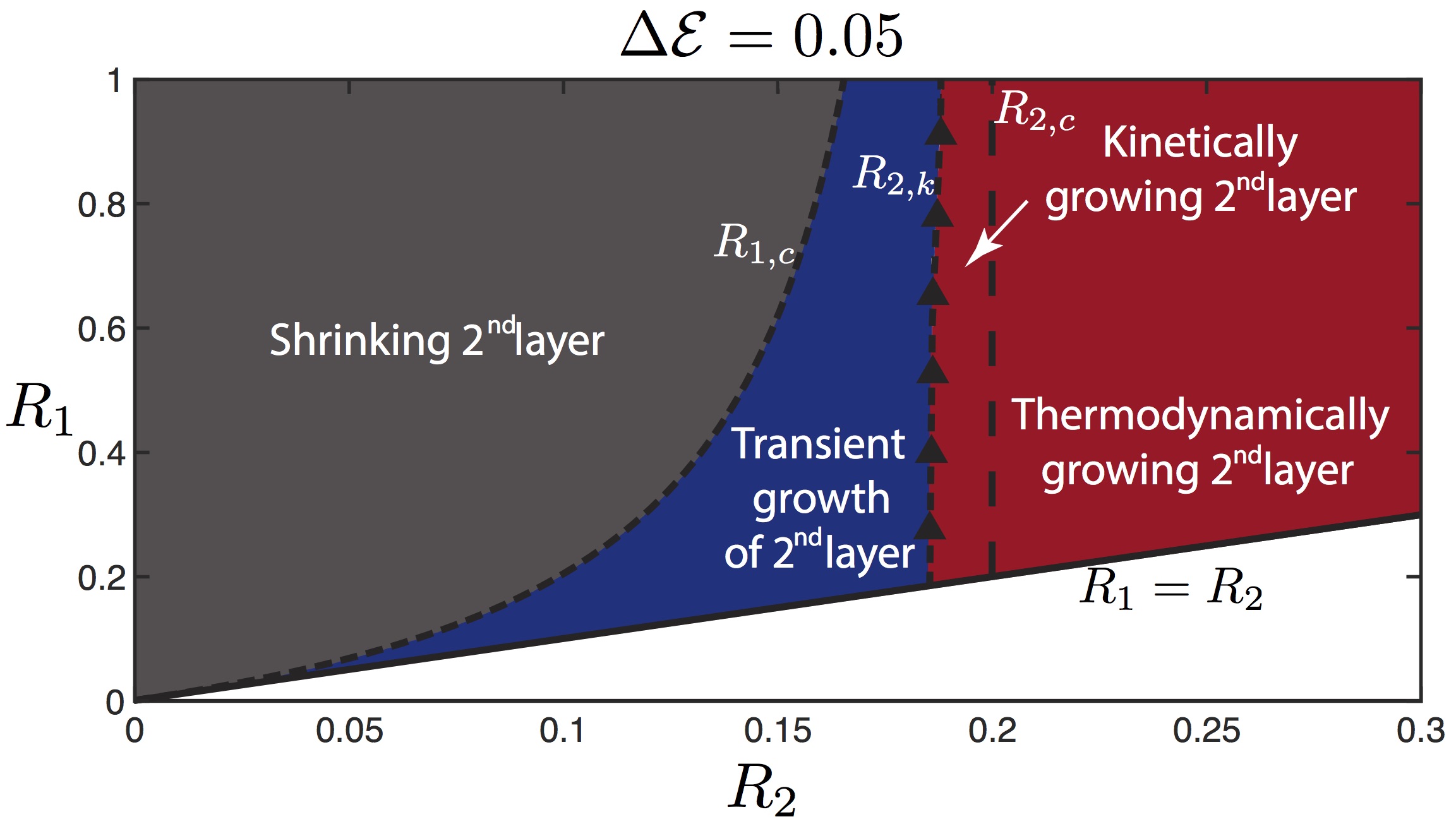}(a)
                 \includegraphics[width=0.8\textwidth]{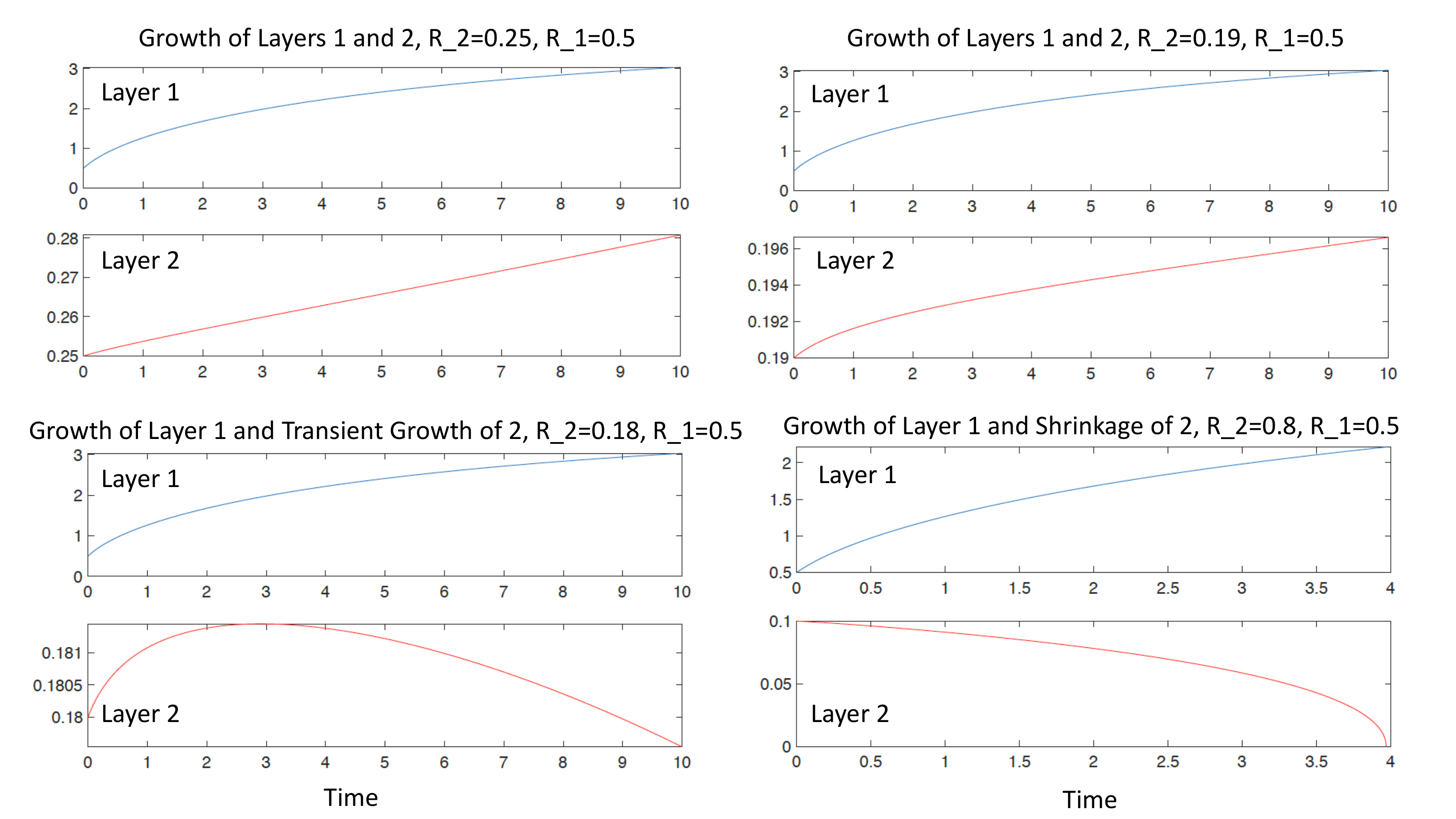}(b)
                       \caption{(a). Morphology diagram, assuming that the layers are circular, showing the dependence of layer 2 growth on the size of layers 1 and 2. In particular, the sign of $v_2$ is shown for different sizes of the layers ($R_1$, $R_2$). When $R_2>R_{2,c}$ layer 2 always grows. When $R_1>R_{1,c}$ layer 2 always shrinks. When $R_{1,c}<R_2<R_{2,k}$ layer 2 grows transiently before shrinking. When $R_{2}>R_{2,k}$ layer 2 grows because $R_2$ increases past $R_{2,c}$ sooner than $R_1$ crosses $R_{1,c}$. See text for details on $R_{1,c}$, $R_{2,k}$ and $R_{2,c}$. (b). Sample trajectories of the layer radii $R_1$ and $R_2$ in time, starting from different initial radii. The parameters are as in Eq. (\ref{parameter-values}) except with $\mathcal{E}_2=1.05$ so that $\Delta\mathcal{E}=0.05$.}
\label{growth phase figure}
\end{figure}

By solving for the radii $R_1$ and $R_2$ numerically and varying the initial radii, we can estimate $R_{2,k}$ numerically and construct a phase diagram for the growth of the 2nd layer. As an example, we fix the parameters $\Delta\mathcal{E}=0.05$, $F_0=0.1$, $\gamma_1=\gamma_2=0.01$, $\rho^*=0.5$, $D_1=1$, $k_2^-=k_1^+=0.5$ and $R_\infty=3.8$. We then vary the initial sizes of the layers $R_1(0)$ and $R_2(0)$, keeping $R_2(0)\ge R_1(0)$. The resulting phase diagram is shown in  Fig. \ref{growth phase figure}(a). Also observe that for $R_2$ in between $R_{1,c}$ and $R_{2,k}$, the 2nd layer grows transiently before shrinking to zero size. Example trajectories of the layer dynamics are shown in Fig. \ref{growth phase figure}(b).

%
%

\section{Reformulation of the vdW-BCF model of multilayer growth using the diffuse domain method}\label{DDM}

To solve the vdW-BCF equations for unconstrained layer geometries, we reformulate the system using the diffuse domain method (DDM). Here, we combine and extend the approaches from \cite{000356209300006} and \cite{Ratz2015} to develop a fully-second order accurate DDM for the vdW-BCF system. We embed the substrate and layer domains into a larger, rectangular domain $\tilde{\Omega}$ and we introduce a diffuse domain function $\varphi$ to mark the locations of the layers and substrate (e.g., approximate atomic height). In particular, $\varphi\approx 0$ in the substrate ($\Omega_0$),  $\varphi\approx 1$ in layer 1 ($\Omega_1$) and $\varphi\approx 2$ in layer 2 ($\Omega_2$). 

\begin{figure}
                 \centering
                \includegraphics[width=0.8\textwidth]{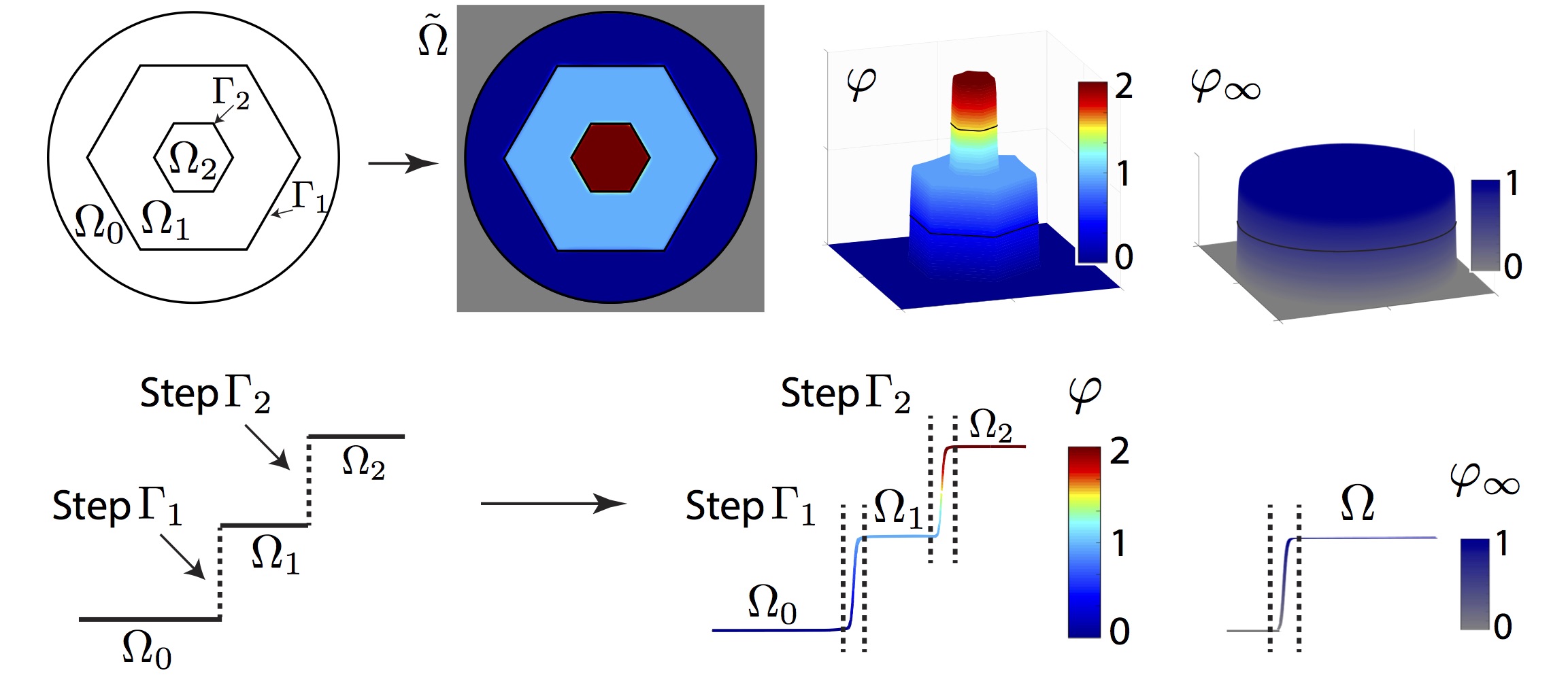}
                       \caption{Schematic of the diffuse domain method. Top: The sharp (physical) interface domain is embedded in a larger, square domain $\tilde\Omega$ where a phase-field functions $\phi$ and $\phi_\infty$ approximate the height of the layers and the characteristic function of the deposition domain, respectively. Bottom: A slice across the sharp interface domain and slices of the phase-field functions $\varphi$ and $\varphi_\infty$.}
\label{diffuse domain method}
\end{figure}

In order to facilitate comparisons with theory from the previous section, we assume that the outer boundary of the substrate is circular and so we introduce another diffuse domain function $\varphi_\infty$ to identify the deposition domain $\Omega=\Omega_0\cup\Omega_1\cup\Omega_2$, where $\varphi_\infty\approx 1$, within the larger domain $\tilde\Omega$. See Fig. \ref{diffuse domain method}(a). The diffuse domain variables change rapidly but smoothly across the boundaries (e.g., steps) as shown in Fig. \ref{diffuse domain method}(b). The width of these narrow transition layers is $\approx \epsilon$, a small parameter. The boundaries of the substrate and layers 1 and 2 correspond to $\varphi\approx 0.5$ and $\varphi\approx 1.5$, respectively. The kinetic boundary conditions are incorporated via source terms and the dynamics of the layers are captured by evolving the diffuse domain function $\varphi$. In addition, we follow \cite{Ratz2015} and solve only two adatom diffusion equations in the extended domain $\tilde\Omega$. A brief description of the derivation and an asymptotic analysis of the vdW-BCF-DDM, which demonstrates that the vdW-BCF-DDM system approximates the sharp interface vdW-BCF model to $O(\epsilon^2)$, are given in \ref{diffuse domain method details}. Here, we present only the resulting equations:
\begin{eqnarray}
\fl (\varphi_\infty H_{0}(\varphi)\rho_{0}^\epsilon)_{t}={{\bf{\nabla}}}\cdot\big (\varphi_{\infty} H_{0}(\varphi)D_0(\varphi){{\bf{\nabla}}}\rho_{0}^\epsilon\big)+\varphi_\infty H_{0}(\varphi)F_{0}(\varphi)\nonumber\\
 -\varphi_{\infty}H_{0}(\varphi)\tau_{d}^{-1} \rho_{0}^\epsilon-\varphi_{\infty}|{\bf{\nabla}}\varphi| k_0(\varphi)\bigg(\rho_{0}^\epsilon-\rho^{*}(\mathcal{E}(\varphi)+\epsilon^{-1}\gamma(\varphi) \mu)\bigg),\label{p1222}\\
\fl (\varphi_{\infty}H_{1}(\varphi)\rho_{1}^\epsilon)_{t}={\bf{\nabla}}\cdot\big (\varphi_{\infty} H_{1}(\varphi)D_{1} {\bf{\nabla}}\rho_{1}^\epsilon\big)+\varphi_{\infty}H_{1}(\varphi) F_{1}\nonumber\\
- \varphi_{\infty}H_{1}(\varphi) \tau_{d}^{-1} \rho_{1}^\epsilon-\varphi_{\infty}|{\bf{\nabla}}\varphi| k_1(\varphi)\bigg(\rho_{1}^\epsilon-\rho^{*}(\mathcal{E}(\varphi)+\epsilon^{-1} \gamma(\varphi) \mu)\bigg),\label{p2222}
\end{eqnarray}
where the kinetic boundary conditions (\ref{BCF-bc11})-(\ref{BCF-bc41}) are modeled by the extra source terms containing $|{\bf{\nabla}}\varphi|$, which approximates the surface delta function. Eq. (\ref{p1222}) models the adatom diffusion equations on the substrate and layer 2, e.g. $\rho_0^\epsilon$ approximates the adatom concentration on both the substrate, where $\varphi\approx 0$, and layer 2, where $\varphi\approx 2$.
Eq. (\ref{p2222}) models adatom diffusion on layer 1 and $\rho_1^\epsilon$ is the corresponding approximate adatom concentration. For simplicity, we have assumed $\tau_{d,i}=\tau_d$. The functions $H_0$, $H_1$  are extended approximate characteristic functions of the layer domains and substrate. In particular, $H_0$ is the approximate characteristic function of the substrate and layer 2:
\begin{eqnarray}
H_{0}(\varphi)=\cases{1-\varphi&for $\varphi \in [0,1]$,\\
\varphi-1&for $\varphi \in (1,2]$,\\}
\end{eqnarray}
and $H_1$ is the approximate characteristic function of layer 1:
\begin{eqnarray}
H_{1}(\varphi)=\cases{\varphi&for $\varphi \in [0,1]$,\\
2-\varphi&for $\varphi \in (1,2]$.\\}
\end{eqnarray}
Further, the flux $F_0(\varphi)$ corresponds to the flux on the substrate
\begin{eqnarray}
F_{0}(\varphi)=\cases{F_{0}&for $\varphi < \epsilon$,\\
0&for $\varphi  \in [\epsilon,2]$,\\}
\end{eqnarray}
and  $D_0(\varphi)$ corresponds to the adatom diffusion coefficients on the substrate and layer 2
\begin{eqnarray}
D_{0}(\varphi)=\cases{D_{0}&for $\varphi \in [0,1]$,\\
D_{2}&for $\varphi \in (1,2]$.\\}
\end{eqnarray}
Analogously, the extended vdW energies and kinetic attachment rates are defined as
\begin{eqnarray}
\mathcal{E}(\varphi)=\cases{-\mathcal{E}_{1}&for $\varphi \in [0,1]$,\\
-\mathcal{E}_{2}+\mathcal{E}_{1}&for $\varphi \in (1,2]$,\\}
\end{eqnarray}
and
\begin{eqnarray}
k_0(\varphi)=\cases{k_{1}^{-}&for $\varphi \in [0,1]$,\\
k^{+}_{2}&for $\varphi \in (1,2]$,\\}
\end{eqnarray}
\begin{eqnarray}
k_1(\varphi)=\cases{k_{1}^{+}&for $\varphi \in [0,1]$,\\
k^{-}_{2}&for $\varphi \in (1,2]$.\\}
\end{eqnarray}

\noindent
The evolution of the layers is implicitly captured by evolving $\varphi$:
\begin{eqnarray}
\fl \partial_{t}\varphi=|{\bf{\nabla}}\varphi|\bigg(k_{0}(\varphi)\big(\rho_{0}^\epsilon-\rho^{*}(\mathcal{E}(\varphi)+\epsilon^{-1}\gamma(\varphi)\mu)\big)
+k_{1}(\varphi)\big(\rho_{1}^\epsilon-\rho^{*}(\mathcal{E}(\varphi)+\epsilon^{-1}\gamma(\varphi)\mu)\big)\bigg) \nonumber\\
+ \epsilon^{-2}\beta{\bf{\nabla}} \cdot (G(\varphi){\bf{\nabla}}\mu ),\label{phase222}\\
\fl \mu=-\epsilon^{2} \Delta\varphi+B'(\varphi),\label{chem222}
\end{eqnarray}	
where the right hand side of Eq. (\ref{phase222}) models the normal velocity from Eq. (\ref{normal-vel-multi}). Note that since the outer boundary of the substrate does not change we do not need to pose an evolution equation for $\varphi_0$. In Eqs. (\ref{phase222}) and (\ref{chem222}), $G(\varphi)=2B(\varphi)$ is an extended double well potential:
\begin{eqnarray}
B(\varphi)=\cases{18\varphi^2(\varphi-1)^2&for $\varphi \in [0,1]$,\\
18(\varphi-1)^2(\varphi-2)^2&for $\varphi \in (1,2]$.\\}
\label{G form}
\end{eqnarray}
As shown in \ref{diffuse domain method details}, and confirmed by our numerical results in the next section, the vdW-BCF-DDM system is second order accurate with respect to the interface thickness $\epsilon$. Moreover, our diffuse interface model can be extended to simulate the more nonlinear model derived in \ref{model-derivation} and to simulate an arbitrary number of vertically-stacked layers (see \ref{arbitrary layers}).

Finally, at the boundary of the larger domain $\partial\tilde{\Omega}$, we take the conditions
\begin{eqnarray}
{\bf{\nabla}}\rho_{0}^\epsilon\cdot {\bf{n}}={\bf{\nabla}}\rho_{1}^\epsilon\cdot {\bf{n}}={\bf{\nabla}}\varphi\cdot {\bf{n}}={\bf{\nabla}}\mu\cdot {\bf{n}}=0.
\label{far field bc}
\end{eqnarray}
The model is insensitive, however, to the choice of boundary conditions on $\partial\tilde\Omega$.

\section{Numerical Results}\label{NumRes}
To solve the vdW-BCF-DDM system (\ref{p1222})-(\ref{far field bc}) numerically, we develop a mass-conservative, semi-implicit, second-order accurate, adaptive finite-difference method using Crank-Nicholson discretization in time and centered differences in space, by extending our previous work, e.g. \cite{BSAM20}. To solve the nonlinear discrete system at the implicit time level, we use a full approximation storage (FAS) nonlinear multigrid method. Block-structured adaptive mesh refinement is utilized to efficiently discretize the system. The details of the method are provided in \ref{app-methods}.

We begin by considering the isotropic, quasi-steady case so we may compare our numerical results to the analytical solutions presented in Sec. \ref{model-analysis} to validate the accuracy of our approximations. We then consider time-dependent diffusion and anisotropic edge energies and kinetic coefficients. We perform parametric studies to determine the effect of parameters on the growth and morphologies of the layers.

\subsection{Quasi-stationary dynamics}\label{quasi-steady}

We consider the same set up as in Sec. \ref{model-analysis}. Initially, two layers are centered at the origin with different radii $R_{1}$ and $R_{2}$ and the edge energy and kinetic coefficients are isotropic. The two islands are bounded by a larger circular substrate with radius $R_{0}$. The initial condition for the diffuse domain variable is
\begin{eqnarray}
\varphi (x,0) &= \frac{1}{2}\left(1 - {\rm tanh} \left(\frac{ 3(x-R_{1})}{\epsilon}\right)\right)+\frac{1}{2}\left(1 - {\rm tanh} \left(\frac{ 3(x-R_{2})}{\epsilon}\right)\right),\label{initial-condition-steps}
\end{eqnarray}
such that $\varphi\approx 1$ approximates layer 1 and $\varphi \approx 2$ approximates layer 2. We take
\begin{eqnarray}
\varphi_{\infty} (x) &= \frac{1}{2}\left(1 - {\rm tanh} \left(\frac{ 3(x-R_{0})}{\epsilon}\right)\right),\label{initial-condition-substrate}
\end{eqnarray}
which corresponds to the region containing the substrate and the two layers where deposition and growth take place. The parameter $\epsilon$ is the thickness of the layer and substrate boundaries. The initial radii of the layers are $R_1(0)=1.2$ and $R_2(0)=0.6$. The outer radius of the substrate is $R_\infty=1.8$. The physical parameters are taken to be
\begin{eqnarray}
\fl k^{\pm}_{1} = k^{\pm}_{2} = 1,~~ \rho^* = 0.01, ~~\gamma_{1} =\gamma_{2} = 1, ~~D_{0} =D_{1} =D_{2} = 1, ~~F_{0} =F_{1} =F_{2} = 2, \nonumber\\
\fl\tau^{-1}_{d} = 0,~~~\beta = 0,~~\mathcal{E}_{1}=-1,~~\mathcal{E}_{2}=-2.\label{paramterss}
\end{eqnarray}
The computations are carried out on a square domain $[-2,2]\times [-2,2]$. A 4-level adaptive mesh is employed, which consists of a root level with mesh size $h_0$ and three refinement levels above it so that the finest mesh size $h_{3}=h_0/8$. In order to test the convergence rate corresponding to different values of $\epsilon$, we refine the root level grid size $h_{0}$ and $\epsilon$ together, and hence all the finer level grid sizes $h_{1}$, $h_{2}$ and $h_{3}$ are refined as well. In particular, we set $h_{3}= \frac{\epsilon}{6.4}$. The mesh is refined according to values of $|{\bf{\nabla}} \varphi|+|{\bf{\nabla}} {\varphi}_{0}|$ over the entire domain (see \ref{app-methods}). The time step is taken to be $\Delta t =\frac{\epsilon}{0.8}\times 10^{-4}$ to ensure that the time errors are small compared to spatial errors; the method is stable (and accurate) for larger time steps.

Five different values of $\epsilon$ are used for the convergence test, namely, $\epsilon_{1} = 0.8$, $\epsilon_{2} = 0.4$, $\epsilon_{3} = 0.2$, $\epsilon_{4} = 0.1$ and $\epsilon_{5} = 0.05$. 
The difference between the analytical solutions and our numerical results are computed using the following metrics:
\begin{eqnarray}
E^{(2)}_{\epsilon,\rho_k} = \frac{|| \varphi(\rho_k^\epsilon - {\rho}_{k})||_{\ell _{2}} }{||\varphi {\rho}_k ||_{\ell _{2}}} ~~{\rm and}~~E^{(\infty)}_{\epsilon,\rho_k} = \frac{|| \varphi(\rho^\epsilon_{k} - {\rho}_{k})||_{\ell _{\infty}} }{||\varphi {\rho}_{k}||_{\ell _{\infty}}}, 
\end{eqnarray}
where $k=0$ denotes the substrate and $k=1$, 2 denote the layers. The convergence rate is obtained by $r_{i-1} = \ln{E^{(\cdot)}_{\epsilon_{i},\rho_k}/ E^{(\cdot)}_{\epsilon_{i-1},\rho_k}}$, where $\epsilon_i$ and $\epsilon_{i-1}$ represent consecutive values of $\epsilon$. The horizontal slices of the adatom concentrations $\rho^\epsilon_{k}$  for different $\epsilon$ together with the analytical solution are shown at time $t=0.1$ in Fig. \ref{ana-vs-numerical}(a). We can observe that the numerical results approach the analytical solution as $\epsilon$ decreases. The corresponding errors and rates of convergence are presented in Table \ref{51l2andlinf}, which indicates that the numerical method converges to the analytic solution with an overall second order convergence rate in both the $\ell_{2}$ and $\ell_{\infty}$ norms, as predicted by the asymptotic analysis in \ref{diffuse domain method details}.

\begin{table}[h]
\caption{Convergence test for the adatom concentrations $\rho_{2}$, $\rho_{1}$ and $\rho_{0}$ under quasi-steady dynamics from \S \ref{quasi-steady}.}
\label{51l2andlinf}

\begin{indented}
\lineup
\item[]
	\begin{tabular}{c||c|c||c|c||c|c}
	\br                              
 t=0.10& $\ell_{2}$\\
	\hline
$\epsilon$     & $E^{(2)}_{\epsilon,\rho_{2}}$ & rate& $E^{(2)}_{\epsilon,\rho_{1}}$ & rate& $E^{(2)}_{\epsilon,\rho_{0}}$ & rate
	\\
	\hline
0.8 &  5.252$\times 10^{-2}$& ---& 5.143$\times 10^{-2}$& ---&  2.095$\times 10^{-1}$& ---
	\\
	\hline
0.4 &  1.514$\times 10^{-2}$& 1.80&  2.020$\times 10^{-2}$& 1.35&  6.557$\times 10^{-2}$& 1.66
	\\
	\hline
0.2 &  3.229$\times 10^{-3}$& 2.23&  3.858$\times 10^{-3}$& 2.40& 1.529$\times 10^{-2}$& 2.10
	\\
	\hline
0.1 &  8.198$\times 10^{-4}$& 1.98&  9.945$\times 10^{-4}$& 1.96&  4.198$\times 10^{-3}$& 1.87\\
	\hline
0.05 &  1.801$\times 10^{-4}$& 2.12& 2.411$\times 10^{-4}$& 2.04&  1.001$\times 10^{-3}$& 2.07
	\\
	\hline
t=0.10& $\ell_{\infty}$\\
	\hline
$\epsilon$     & $E^{(\infty)}_{\epsilon,\rho_{2}}$ & rate& $E^{(\infty)}_{\epsilon,\rho_{1}}$ & rate& $E^{(\infty)}_{\epsilon,\rho_{0}}$ & rate
	\\
	\hline
0.8 &  6.132$\times 10^{-2}$& ---& 5.812$\times 10^{-2}$& ---&  3.271$\times 10^{-1}$& ---
	\\
	\hline
0.4 &  1.914$\times 10^{-2}$& 1.68&  2.360$\times 10^{-2}$& 1.30&  1.214$\times 10^{-1}$& 1.43
	\\
	\hline
0.2 &  5.058$\times 10^{-3}$& 1.92&  5.429$\times 10^{-3}$& 2.12& 3.368$\times 10^{-2}$& 1.85
	\\
	\hline
0.1 &  1.453$\times 10^{-3}$& 1.80&  1.496$\times 10^{-3}$& 1.86&  1.058$\times 10^{-2}$& 1.67\\
	\hline
0.05 &  3.946$\times 10^{-4}$& 1.88& 4.007$\times 10^{-4}$& 1.90&  2.996$\times 10^{-3}$& 1.82
\\
\br
\end{tabular}
\end{indented}
	\end{table}

\subsection{Fully time-dependent case}\label{double-fully-time}

\begin{figure}
                  \centering
                \includegraphics[width=0.7\textwidth]{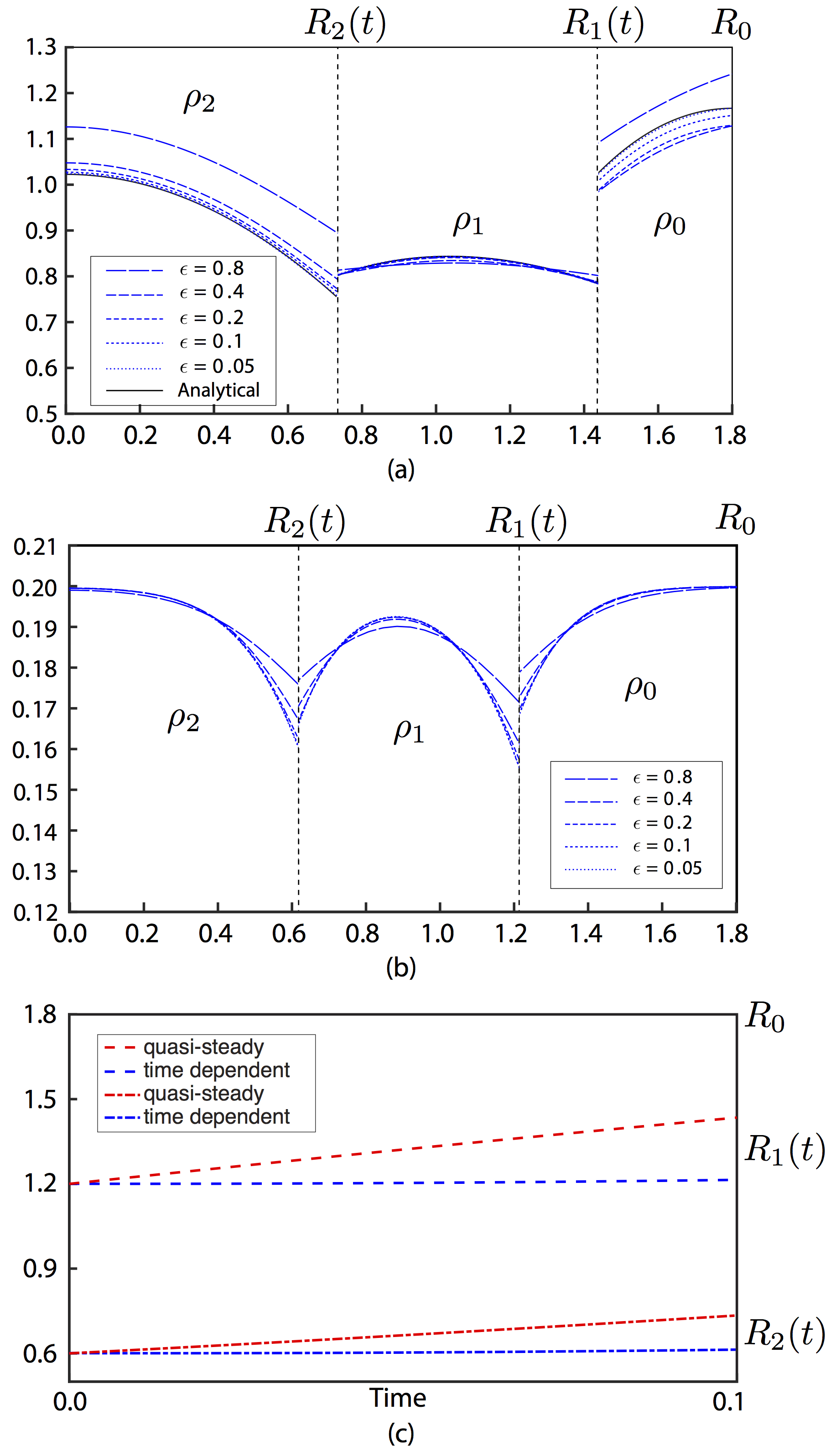}
                       \caption{Comparisons between the numerical results and analytical solution at time $t=0.1$. In (a), the quasi-steady dynamics are shown from  \S \ref{quasi-steady}. In (b), the fully-time dependent dynamics are shown from \S \ref{double-fully-time}. The dashed and dotted lines represent the horizontal slices of densities $\rho_{2}$, $\rho_{1}$ and $\rho_{0}$ at different $\epsilon$, as labeled. In (a), the black solid lines give the analytical solution. The radii of the layers are shown as a function of time in (c). The adatom concentrations and gradients are larger in the quasi-steady case, which give rise to faster dynamics in the quasi-steady case.}\label{ana-vs-numerical}
\end{figure}

Next, we include the time derivatives in the adatom diffusion equations. The physical parameters, the computational domain and the numerical parameters are the same as in the previous section. Since we do not have an analytic solution in this case, we compare the results obtained using different $\epsilon$ (and hence $h_0$) with each other. The horizontal slices of the adatom concentrations $\rho^\epsilon_2$, $\rho^\epsilon_1$ and $\rho^\epsilon_0$ are shown at time $t=0.1$ in Fig. \ref{ana-vs-numerical}(b). Compared to the quasi-steady case, the adatom concentrations in each layer are smaller and there is less variation across the layers. Correspondingly, the layers do not move as rapidly in the time-dependent case with the first layer growing more slowly than the second, compared to the quasi-steady case (Fig. \ref{ana-vs-numerical}(c)).  Fig. \ref{ana-vs-numerical}(b) also shows that the results converge as $\epsilon$ is decreased. To estimate the accuracy and quantify the rate of convergence, we define the consecutive errors  as
\begin{eqnarray}
\fl E^{(2)}_{\epsilon_{i-1},\epsilon_i,\rho_{2}}&= || \varphi_{2} \left(\rho^{\epsilon_{i-1}}_{2} - \rho^{\epsilon_{i}}_{2}\right)||_{\ell_2},\hspace{5mm}E^{(\infty)}_{\epsilon_{i-1},\epsilon_i,\rho_{2}}= || \varphi_{2} (\rho^{\epsilon_{i-1}}_{2} - \rho^{\epsilon_{i}}_ {2})||_{\ell_{\infty}}\nonumber\\
\fl E^{(2)}_{\epsilon_{i-1},\epsilon_i,\rho_{1}}&= || \varphi_{1} (\rho^{\epsilon_{i-1}}_{1} - \rho^{\epsilon_{i}}_ {1})||_{\ell_2},\hspace{6mm}E^{(\infty)}_{\epsilon_{i-1},\epsilon_i,\rho_{1}}= || \varphi_{1} (\rho^{\epsilon_{i-1}}_{1} - \rho^{\epsilon_{i}}_ {1})||_{\ell_{\infty}},\nonumber\\
\fl E^{(2)}_{\epsilon_{i-1},\epsilon_i,\rho_{0}}&= || \varphi_{\infty} (\rho^{\epsilon_{i-1}}_{0} - \rho^{\epsilon_{i}}_ {0})||_{\ell_2},\hspace{6mm}E^{(\infty)}_{\epsilon_{i-1},\epsilon_i,\rho_{0}}= || \varphi_{\infty} (\rho^{\epsilon_{i-1}}_{0} - \rho^{\epsilon_{i}}_ {0})||_{\ell_{\infty}},
\label{consecutive errors}
\end{eqnarray}
where the $\varphi_j$, with $j=0$, 1 and 2 are the approximate characteristic functions on the substrate, layer 1 and layer 2 respectively. They are defined as
\begin{eqnarray}
\varphi_2=\cases{ \varphi- 1 &for $\varphi \in (1,2]$,\\
0&for $\varphi \in [0,1]$,\\}
\end{eqnarray}
\begin{eqnarray}
\varphi_1=\sqrt{(1-\sqrt{(\varphi-1)^2})^2},
\end{eqnarray}
\begin{eqnarray}
\varphi_\infty=\cases{ 0 &for $\varphi \in (1,2]$,\\
\varphi_\infty\left(1- \varphi\right)&for $\varphi \in [0,1]$,\\}
\end{eqnarray}
and these functions are evaluated at $\epsilon=\epsilon_i$. The errors and rates of convergence, which are calculated from the consecutive errors at time $t=0.1$ in an analogous way as in the previous section, are presented in Table \ref{tab112}. As in the quasi-steady case, we observe that the results converge with second order accuracy in both the $\ell_{2}$ and $\ell_{\infty}$ norms.

\begin{table}[h]
\caption{Convergence test for concentrations $\rho_{2}$, $\rho_{1}$ and $\rho_{0}$ under the fully time-dependent dynamics in \S \ref{double-fully-time}.}
\label{tab112}
\begin{indented}
\lineup
\item[]
	\begin{tabular}{c||c|c||c|c||c|c}
\br
 t=0.10& $\ell_{2}$\\
	\hline
$\epsilon$     & $E^{(2)}_{\epsilon_{i-1},\epsilon_i,\rho_{2}}$ & rate& $E^{(2)}_{\epsilon_{i-1},\epsilon_i,\rho_{1}}$ & rate& $E^{(2)}_{\epsilon_{i-1},\epsilon_i,\rho_{0}}$ & rate
	\\
	\hline
0.4 &  3.978$\times 10^{-3}$& ---& 8.066$\times 10^{-3}$& ---&  5.648$\times 10^{-3}$& ---
	\\
	\hline
0.2 &  2.315$\times 10^{-3}$& 0.78&  3.299$\times 10^{-3}$& 1.29&  2.224$\times 10^{-3}$& 1.35
	\\
	\hline
0.1 &  6.977$\times 10^{-4}$& 1.73&  9.287$\times 10^{-4}$& 1.83& 5.779$\times 10^{-4}$& 1.94
	\\
	\hline
0.05 &  1.781$\times 10^{-4}$& 1.97&  2.258$\times 10^{-4}$& 2.04&  1.239$\times 10^{-4}$& 2.22\\
	\hline
 t=0.10& $\ell_{\infty}$\\
	\hline
$\epsilon$     & $E^{(\infty)}_{\epsilon_{i-1},\epsilon_i,\rho_{2}}$ & rate& $E^{(\infty)}_{\epsilon_{i-1},\epsilon_i,\rho_{1}}$ & rate& $E^{(\infty)}_{\epsilon_{i-1},\epsilon_i,\rho_{0}}$ & rate
	\\
	\hline
0.4 &  4.316$\times 10^{-3}$& ---& 4.167$\times 10^{-3}$& ---&  3.241$\times 10^{-3}$& ---
	\\
	\hline
0.2 &  2.521$\times 10^{-3}$& 0.78&  2.299$\times 10^{-3}$& 0.86&  1.816$\times 10^{-3}$& 0.84
	\\
	\hline
0.1 &  9.029$\times 10^{-4}$& 1.48&  8.187$\times 10^{-4}$& 1.49& 6.444$\times 10^{-4}$& 1.49
	\\
	\hline
0.05 &  2.606$\times 10^{-4}$& 1.80&  2.368$\times 10^{-4}$& 1.80&  1.892$\times 10^{-4}$& 1.77\\
\br
\end{tabular}
\end{indented}
	\end{table} 	

\subsection{Anisotropic dynamics}
\label{anisotropic dynamics}
We now consider the case in which the edge energies and kinetic coefficients are anisotropic:
\begin{eqnarray}
k_{0}(\varphi,\theta)=\cases{ k^{+}_{2}\xi_{k}(\theta)&for $\varphi \in (1,2]$,\\
k^{-}_{1}\xi_{k}(\theta)&for $\varphi \in [0,1]$,\\}
\end{eqnarray}
\begin{eqnarray}
k_{1}(\varphi,\theta)=\cases{ k^{-}_{2}\xi_{k}(\theta)&for $\varphi \in (1,2]$,\\
k^{+}_{1}\xi_{k}(\theta)&for $\varphi \in [0,1]$,\\}
\end{eqnarray}
where
\begin{equation}
\xi_{k}(\theta) = 1 - \epsilon_{k,n} {\rm cos }\left(n \left(\theta - \theta_{0}\right) \right),\label{kinetic-coefficients}
\end{equation}
is the kinetic coefficient anisotropy function, $\theta$ is the normal angle (e.g., angle between the normal vector and the $x$-axis), and $\theta_{0}$ is a reference angle which is taken to be $\theta_{0}=\pi/n$.
The edge energies are defined analogously:
\begin{eqnarray}
\gamma(\theta)&=\gamma( \xi_{s}(\theta) + \xi''_{s}(\theta)),\\
\xi_{s}(\theta) &= 1 - \epsilon_{s,n} {\rm cos }(n \theta),
\end{eqnarray}
where $\xi_s$ is the edge energy anisotropy function. The coefficients $\epsilon_{k,n}$ and $\epsilon_{s,n}$ measure the anisotropy strengths. In this paper, we only consider 3-fold ($n=3$) and 6-fold ($n=6$) anisotropies, which reflect the symmetries of $MoS_2$ and graphene multilayers, respectively. The trigonometric functions are calculated  using $\varphi$:
\begin{eqnarray}
{\rm cos}(\theta) = \frac{\varphi_{x}}{\sqrt{\varphi^2_{x}+\varphi^{2}_{y}+\delta}},~~~~{\rm and}~~~~{\rm sin}(\theta) = \frac{\varphi_{y}}{\sqrt{\varphi^2_{x}+\varphi^{2}_{y}+\delta}}~,
\end{eqnarray}
where we introduce a small parameter $\delta=10^{-6}$ to avoid singularities. Then, $\cos(3\theta)$ and $\cos(6\theta)$ can be calculated using the trigonometric identites:
\begin{eqnarray}
\cos(6\theta)&=2\cos^2(3\theta)-1, ~~~~\cos(3\theta)=\cos\theta\cdot\left(2\cos(2\theta)-1\right),\\
\cos(2\theta)&=\cos^2\theta-\sin^2\theta=\frac{\varphi_x^2-\varphi_y^2}{\varphi_x^2+\varphi_y^2+\delta}.
\label{trig formulas}
\end{eqnarray}

\begin{figure}
                 \centering
                \includegraphics[width=1\textwidth]{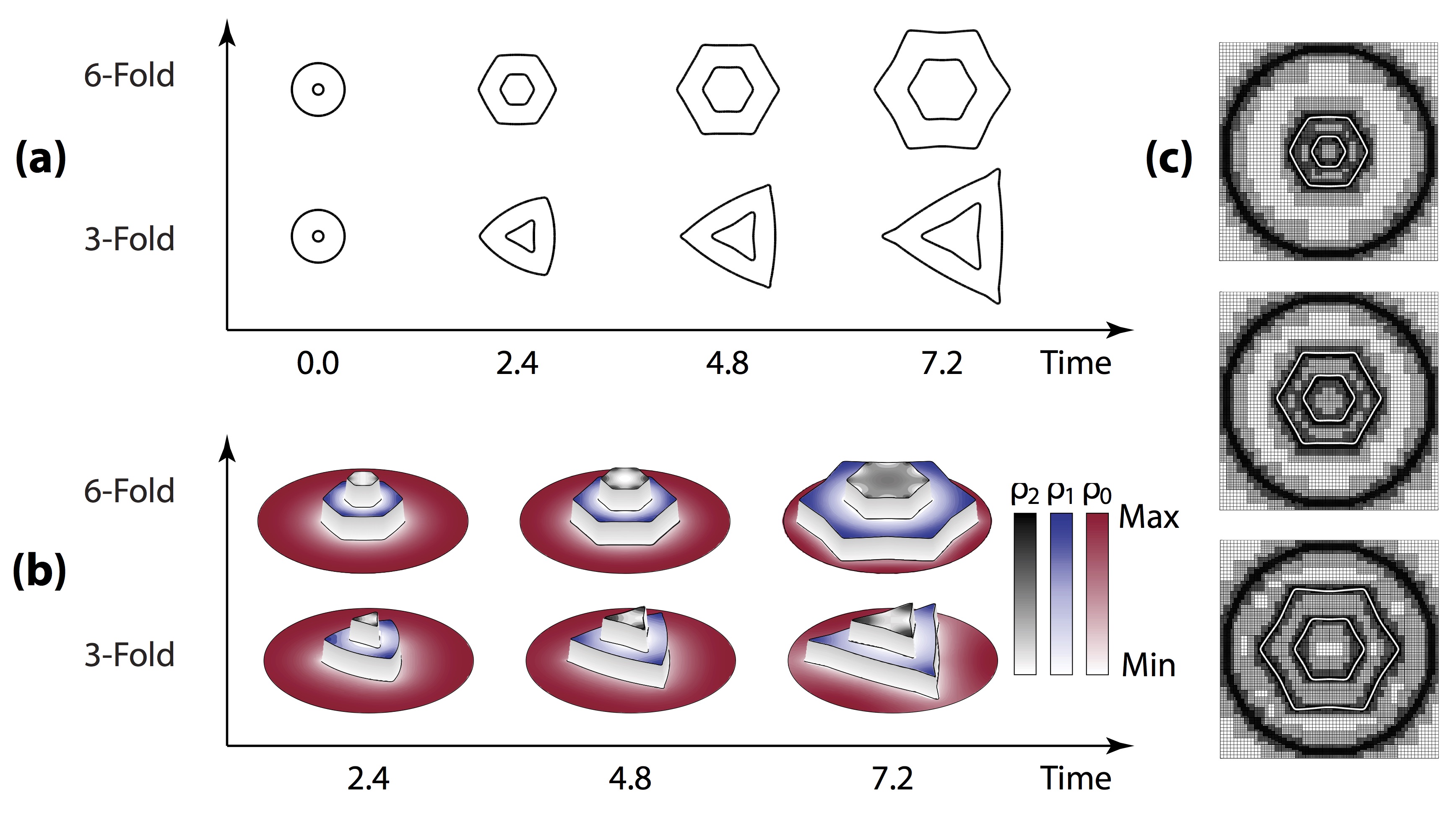}
                       \caption{The quasi-steady dynamics of layers 1 and 2 using $6$-fold and $3$-fold symmetric anisotropic edge energies and kinetic coefficients under conditions for which both layers should grow. See text for parameters. (a). Time evolution of the layer morphologies; (b) Time evolution of the adatom concentrations on the layers. (c). The dynamic adaptive mesh for 6-fold anisotropic layers. As both layers grow, driven by fluxes of the adatoms, negative curvatures develop in both layers in the 6-fold case and in layer 2 in the 3-fold case. The corners of the layers are more affected by surface diffusion in the 3-fold case compared to the 6-fold case.}
\label{aniso-qs}
\end{figure}

We next consider the quasi-steady dynamics of two anisotropic layers. Initially, the layers are taken to be circular with radii $R_1(0)=1.0$ and $R_2(0)=0.2$. The outer boundary of the substrate is $R_\infty=3.8$. The physical parameters are taken to be:
\begin{eqnarray}
\fl k^{\pm}_{1} = k^{\pm}_{2} = 0.5,~~ \rho^*_{1} = 0.5, ~~\gamma_{1} =\gamma_{2} = 0.01, ~~D_{0} =D_{1} =D_{2} = 1, ~~F_{0} = 0.1, \nonumber\\
\fl F_{1} =F_{2} = 0,~~ \tau^{-1}_{d} = 0,~~~\beta = 1.11\times 10^{-5},~~\mathcal{E}_{1}=0.5,~~\mathcal{E}_{2}=1.4.\label{parameter-values}
\end{eqnarray}
Note that unlike the previous examples, the only non-zero flux is on the substrate $F_0$, which as discussed before reflects the assumption that the reactions to produce the attaching species occur only on the substrate surface \cite{Meca2017}.
Note that $\Delta\mathcal{E}=0.4$ and $\gamma_2/R_2(0)-\gamma_1/R_1(0)=0.04$ so that growth would occur under isotropic, quasi-steady dynamics (recall the growth condition in Eq. (\ref{criterion})). The parameters for the anisotropy are set as
\begin{eqnarray}
n=6, ~~\epsilon_{k,n} =0.3,~~\epsilon_{s,n}=0.01,\nonumber\\
n=3, ~~\epsilon_{k,n} =0.7,~~\epsilon_{s,n}=0.01.\label{anisotropic-parameters}
\end{eqnarray}
The morphologies of the growing layers are shown in Fig. \ref{aniso-qs}(a). In both the 6-fold and 3-fold anisotropic cases, layers 1 and 2 grow. In the 6-fold case, the layers are nearly faceted at early times while the corners are smoothed slightly from the surface diffusion. At later times, both layers develop negative curvature. In the 3-fold case, layer 1 evolves to a convex triangular shape at early times while layer 2 develops negative curvature early on. At later times, the corners of layer 1 somewhat elongate with their curvature being set by the surface diffusion coefficient (see Fig. \ref{Interface-comparison-effect-gamma-f}(c)). The corresponding adatom concentrations are shown in Fig. \ref{aniso-qs}(b) where we see the adatoms diffusing toward both layers driving their growth. In Fig.  \ref{aniso-qs}(c), the adaptive mesh is shown for the 6-fold anisotropic case. Observe that there is a fine mesh near the outer boundary of the substrate, which does not change. The mesh near the boundaries of layers 1 and 2 is dynamically refined and and the mesh in the bulk regions is coarsened. In the anisotropic case, we also observe second-order accurate convergence in $\ell_{2}$ and $\ell_{\infty}$, see \ref{convergence_appendix}.

\subsection{Parameter studies}

We next investigate the effects of the physical parameters on the growth of the layers. In particular, we consider the binding energy differences $\Delta\mathcal{E}$, the edge energy $\gamma$ and the surface diffusion $\beta$, flux $F_0$ and the kinetic attachment rates $k_2^-$ and $k_1^+$.  We fix all the other parameters as in Eq.(\ref{parameter-values}) and describe only those parameters that are changed.

\paragraph{Binding energy differences.}

\begin{figure}
                  \centering
                \includegraphics[width=0.9\textwidth]{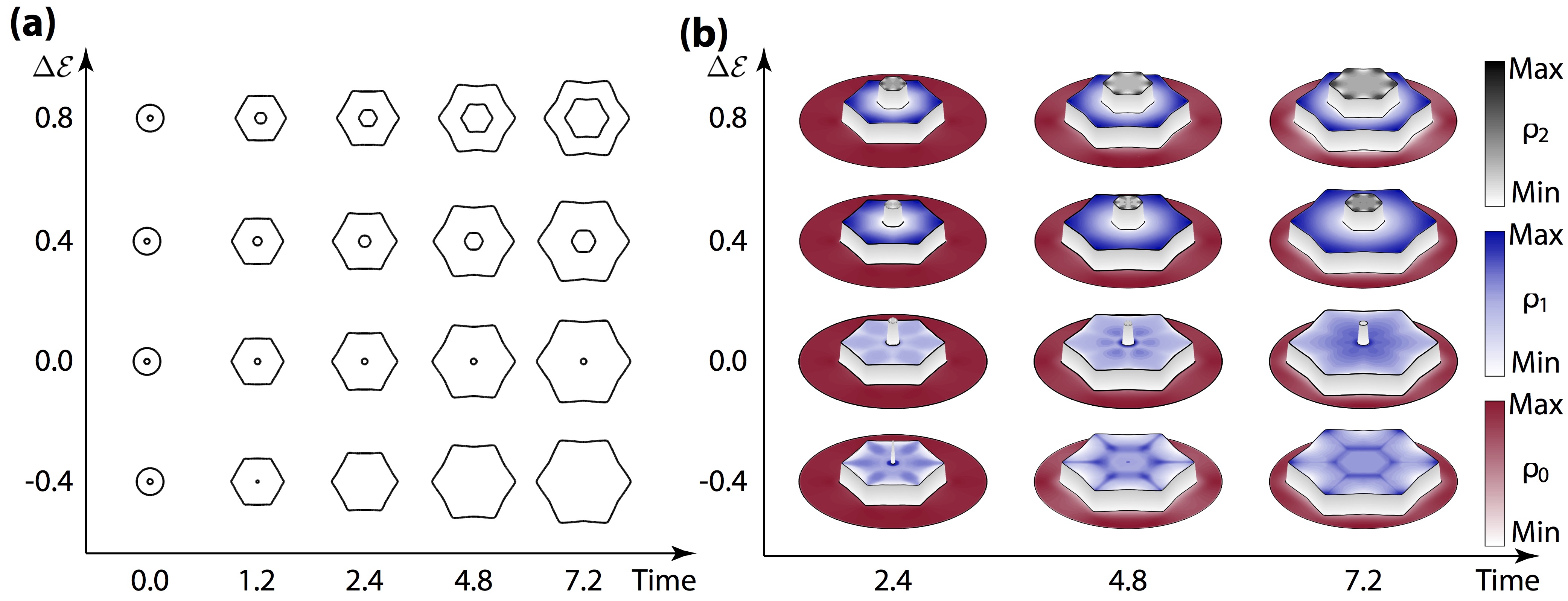}
                       \caption{The effects of binding energy differences $\Delta \mathcal{E}$ for the quasi-steady dynamics of layers 1 and 2 using $6$-fold symmetric anisotropic edge energies and kinetic coefficients. See text for other parameters. (a). Time evolution of  the layer morphologies; (b) Time evolution of the adatom concentrations on the layers. The growth conditions for 6-fold anisotropic layers follow the thermodynamic criterion in Eq. (\ref{criterion}), derived in the isotropic, quasi-steady case (circular layers), that relates $\Delta\mathcal{E}$ and the sizes of the layers. Further, when layer 2 grows, it does so at the expense of layer 1. }\label{Interface-comparison-effect-EE-6fold-quasi-steady}
\end{figure}

We first investigate the effects of $\Delta\mathcal{E}$ on the growth rate of layer 2. The morphologies and adatom concentrations for 6-fold anisotropic layers obtained from the quasi-steady dynamics are shown in Figs. \ref{Interface-comparison-effect-EE-6fold-quasi-steady}(a) and (b), respectively. Consistent with theory (Sec. \ref{model-analysis}), the vertical growth of layer 2 is only preferable when $\Delta \mathcal{E}>0.04$, based on Eqs. (\ref{criterion}) and (\ref{parameter-values}), and that growth rate increases with $\Delta \mathcal{E}$. Further, the growth of layer 2 occurs at the expense of that of layer 1; the size of layer 1 is a decreasing function of $\Delta\mathcal{E}$. In all the cases, layer 1 is nearly faceted at early times, and develops negative curvatures at  late times as layer 1 increases in size. Similar morphologies are observed for layer 2 with negative curvatures occurring when layer 2 is large enough (e.g., $\Delta\mathcal{E}=0.8$). 

\begin{figure}
                  \centering
                \includegraphics[width=0.9\textwidth]{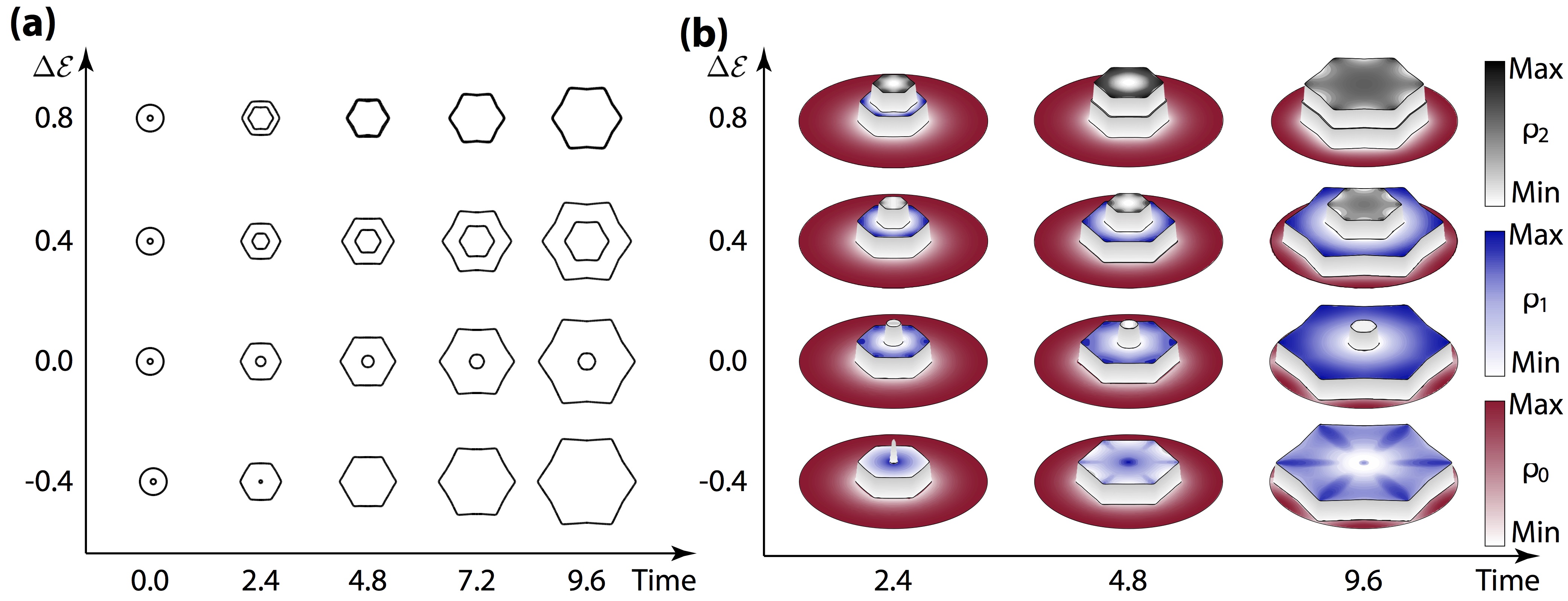}
                       \caption{The effects of binding energy differences $\Delta \mathcal{E}$ for the time-dependent dynamics of layers 1 and 2 using $6$-fold symmetric anisotropic edge energies and kinetic coefficients. (a). Time evolution of  the layer morphologies; (b) Time evolution of the adatom concentrations on the layers. Compared to the quasi-steady case shown in Fig. \ref{Interface-comparison-effect-EE-6fold-quasi-steady}, layer 1 grows more slowly but layer 2 grows more rapidly. In fact, layer 2 grows even when $\Delta\mathcal{E}=0$, in contrast to the quasi-steady case where layer 2 shrinks when $\Delta\mathcal{E}=0$.}\label{Interface-comparison-effect-EE-6fold}
\end{figure}

For comparison, the morphologies for 6-fold anisotropy obtained from the fully time-dependent dynamics are shown in \ref{Interface-comparison-effect-EE-6fold}. Compared to the quasi-steady case, we observe that the growth of layer 1 is significantly slower but that layer 2 actually grows more rapidly. Further, layer 2 grows even when $\Delta\mathcal{E}=0$. This reflects the fact that vertical growth is more favorable when the growth rate of layer 1 is decreased, which is suggested by the theory in Sec. \ref{model-analysis}.

\begin{figure}
                  \centering
                \includegraphics[width=0.9\textwidth]{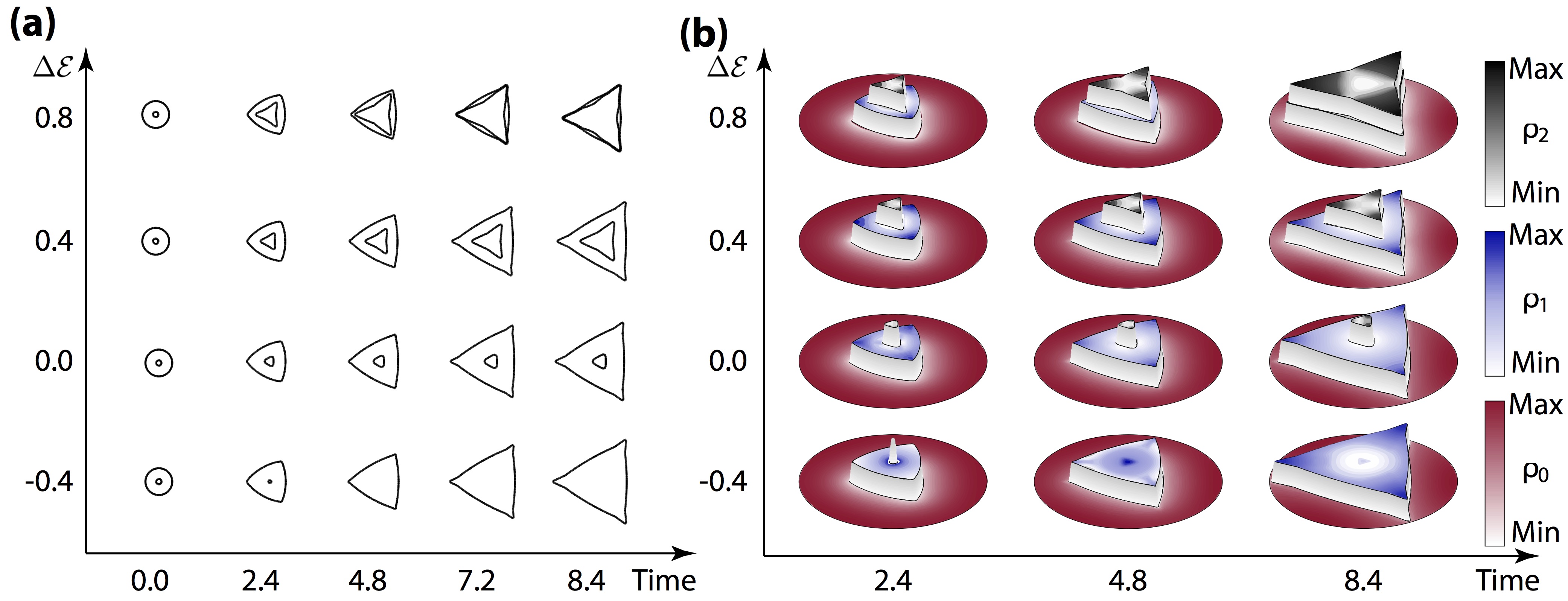}
                       \caption{The effects of binding energy differences $\Delta \mathcal{E}$ for the time-dependent dynamics of layers 1 and 2 using $3$-fold symmetric anisotropic edge energies and kinetic coefficients. (a). Time evolution of  the layer morphologies; (b) Time evolution of the adatom concentrations on the layers. Qualitatively the growth criterion for layer 2 growth is similar to that for the 6-fold, fully time-dependent case shown in Fig. ref{Interface-comparison-effect-EE-6fold}. Quantitatively, the layers grow more rapidly in the 3-fold case. Further, the negative curvature of the sides is more pronounced on layer 2.}\label{Interface-comparison-effect-EE-3fold}
\end{figure}

In Figs. \ref{Interface-comparison-effect-EE-3fold}(a) and (b), the morphologies and adatom concentrations are shown, respectively, for 3-fold anisotropic layers using the fully time-dependent dynamics. Qualitatively, the results are similar to the 6-fold case in Fig. \ref{Interface-comparison-effect-EE-6fold} although we observe that negative curvature occurs first in layer 2 before being manifest in layer 1.

\begin{figure}
                  \centering
                \includegraphics[width=0.9\textwidth]{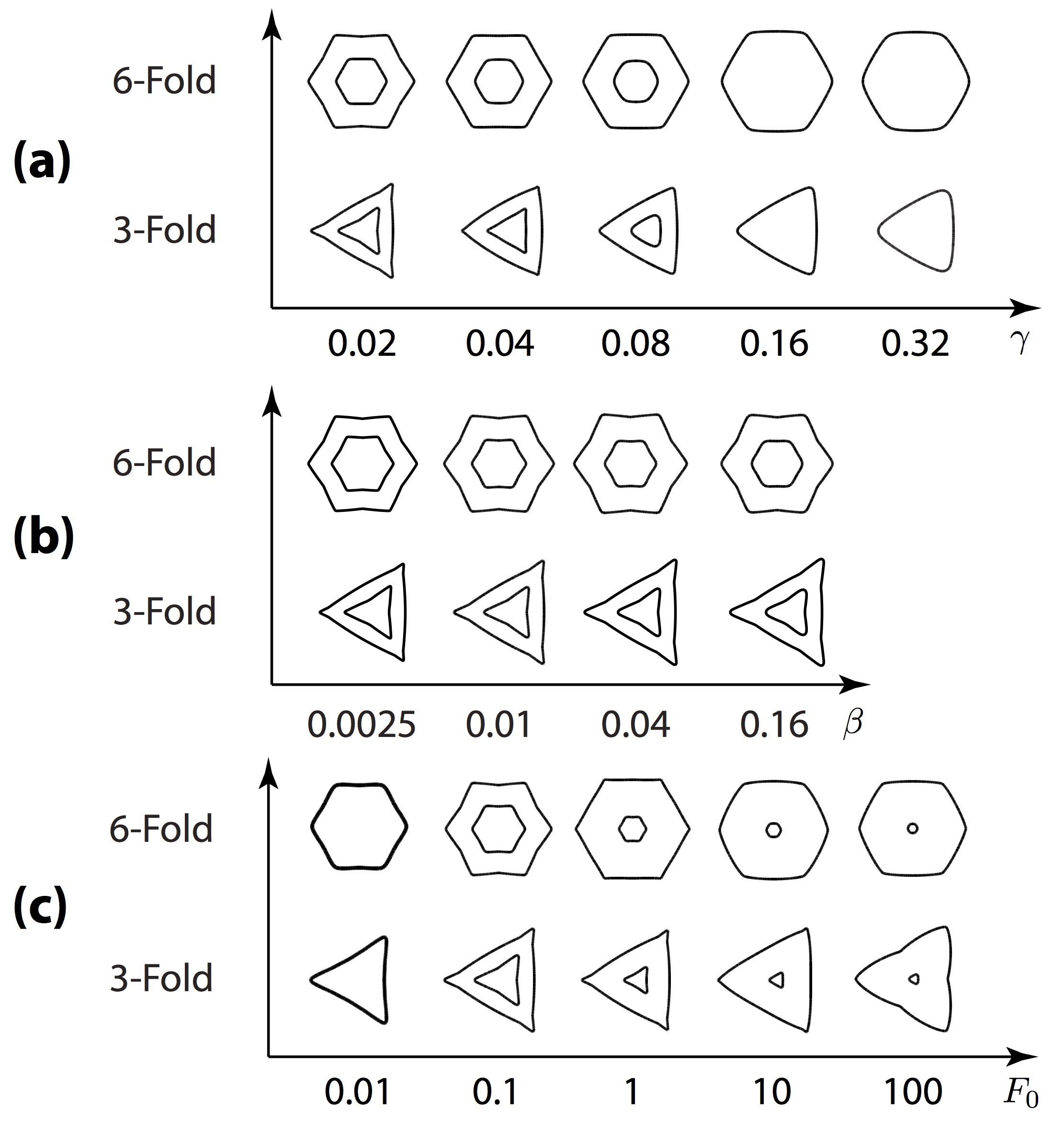}
                       \caption{The morphologies of the 6-fold and 3-fold anisotropic layers at approximately the same sizes for different choices of edge energy strengths $\gamma$ in (a), surface diffusion coefficients $\beta$ in (b) and deposition fluxes $F_0$ in (c). All these parameters inhibit layer 2 growth. In (a) and (b), the 6-fold and 3-fold shapes are shown at times $t=9.6$ and $t=8.4$, respectively. In (c), the times shown for the 6-fold case are ($F_0=0.01$: $t=60$, $F_0=0.1$: $t=9.6$, $F_0=1.0$: $t=2.8$, $F_0=10.0$: $t=0.8$, $F_0=10.0$: $t=0.18$) and for the 3-fold case: ($F_0=0.01$: $t=56$, $F_0=0.1$: $t=8.4$, $F_0=1.0$: $t=2.25$, $F_0=10.0$: $t=0.65$, $F_0=10.0$: $t=0.15$). See text for all the parameters.}
\label{Interface-comparison-effect-gamma-f}
 \end{figure}

\paragraph{Edge energy, surface diffusion and flux.}
In Fig.~\ref{Interface-comparison-effect-gamma-f}(a), we show the effects of edge energy $\gamma$ on the growth of the layers in the fully time-dependent case. In both 6-fold and 3-fold anisotropies, we see that the growth rate of layer 2 decreases as we increase $\gamma$, and the layer 2 even shrinks when $\gamma$ is large enough ($\gamma=0.16$ or larger). The size of layer 1 is also decreased and the layer morphologies are smoother and the negative curvature on the layers disappears as $\gamma$ is increased.

As seen in Fig. \ref{Interface-comparison-effect-gamma-f}(b), surface diffusion also decreases the sizes of layer 2 and smoothens the layer corners although the negative curvature of the layers remains. In the 6-fold anisotropic case, layer 1 is also decreased in size as $\beta$ increases while in the 3-fold anisotropic case, layer 1 is actually a little larger due to the decreased curvature at the vertices.

Next, we examine the effects of the adatom flux $F_{0}$ on the layer dynamics. As shown in Fig.~\ref{Interface-comparison-effect-gamma-f}(c), decreasing the supply of adatoms on the substrate ($F_{0}$) benefits the growth of second layer, which agrees with reported experimental observations for vertical growth of 2D materials (e.g., \cite{Ye2017}). Moreover, in the case of 6-fold anisotropy, we see that both layers develop negative curvatures at small $F_{0}$, but as $F_0$ is increased the shapes become more facetted. Similar features are observed in the 3-fold anisotropic case, except when $F_{0}=100$, where kinks with negative curvature develop at the boundary of layer 1. This feature persists under mesh refinement and seems to be associated with deposition only occurring on the substrate. If adatoms are deposited on all the layers, then layer 1 is convex at an equivalent size.

\paragraph{Kinetic coefficients.}

\begin{figure}
                  \centering
                \includegraphics[width=0.9\textwidth]{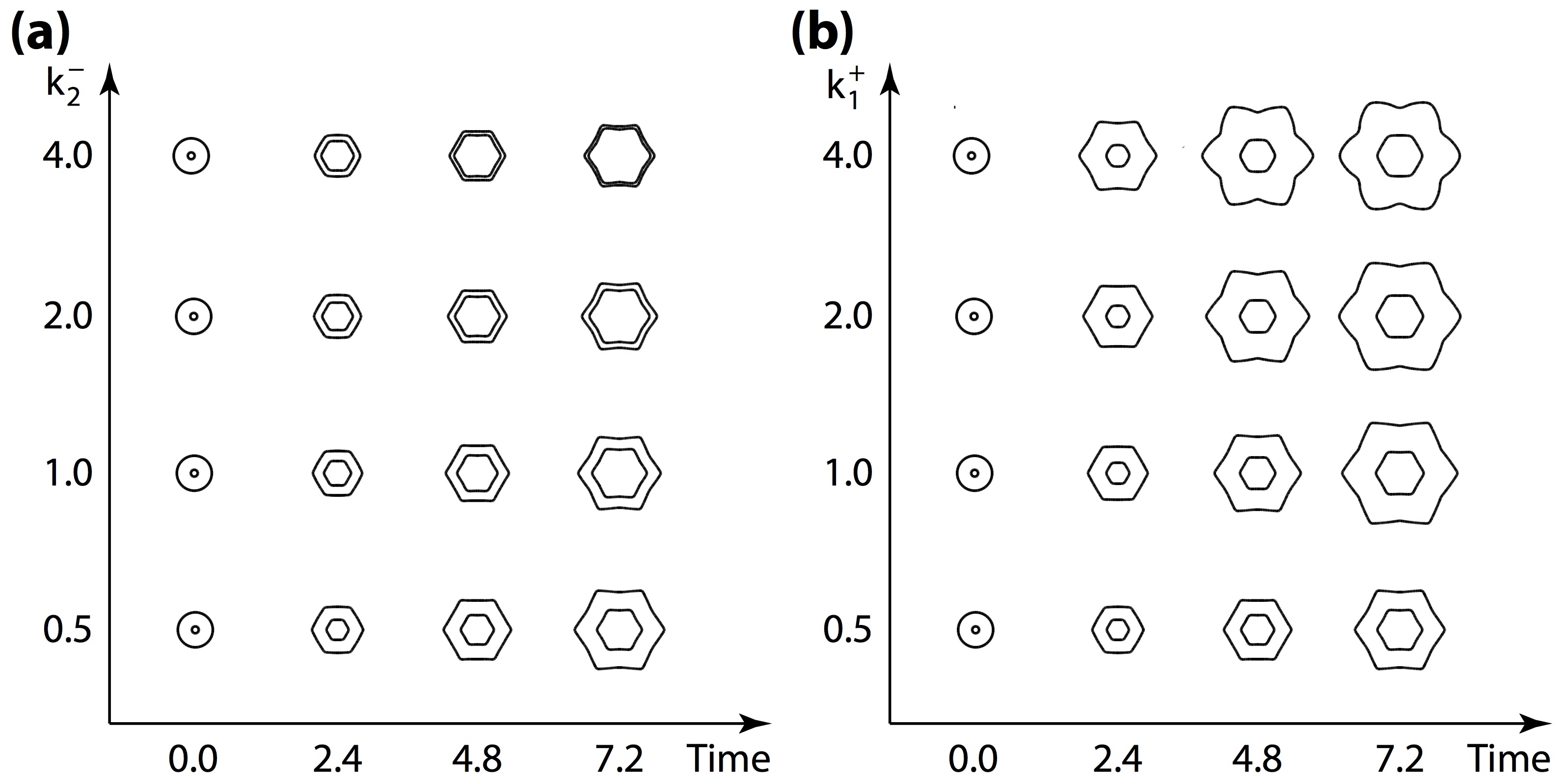}
                       \caption{The effects of kinetic attachment coefficients on the dynamics of 6-fold anisotropic layers 1 and 2. See text for parameters. In (a), only $k^{-}_{2}$ is varied.  In (b), $k_1^+=k_1^-$ are varied together. The kinetic parameter $k_2^-$ promotes layer 2 growth, as predicted by theory. The growth of layer 2 is insensitive to simultaneous changes in $k_1^+$ and $k_1^-$ although layer 1, however, is significantly affected and undergoes a morphological instability. }\label{Interface-comparison-effect-k-6fold}
\end{figure}

In Fig. \ref{Interface-comparison-effect-k-6fold}(a), the kinetic parameter $k_2^-$ is varied from $0.5$ to $4.0$ for layers with 6-fold anisotropies. As predicted by the theory in Sec. \ref{model-analysis}, increasing $k_2^-$ favors the growth of layer 2 at the expense of layer 1. Both layers acquire negative curvature as they grow. In Fig. \ref{Interface-comparison-effect-k-6fold}(b), we take $k_1^+=k_1^-$ and vary this value from $0.5$ to $4.0$. In this case,  the growth of layer 2 is insensitive to these changes, which is surprising because theory suggests that increasing $k_1^+$ increases layer 2 growth (Eq. (\ref{second-layer-velocity-simplified})). The reason for the discrepancy is that a morphological instability occurs on layer 1 that accelerates its growth relative to that of layer 2. Because layer 1 grows faster, this reduces the number of adatoms available for layer 2 growth.

\begin{figure}
                  \centering
                \includegraphics[width=0.9\textwidth]{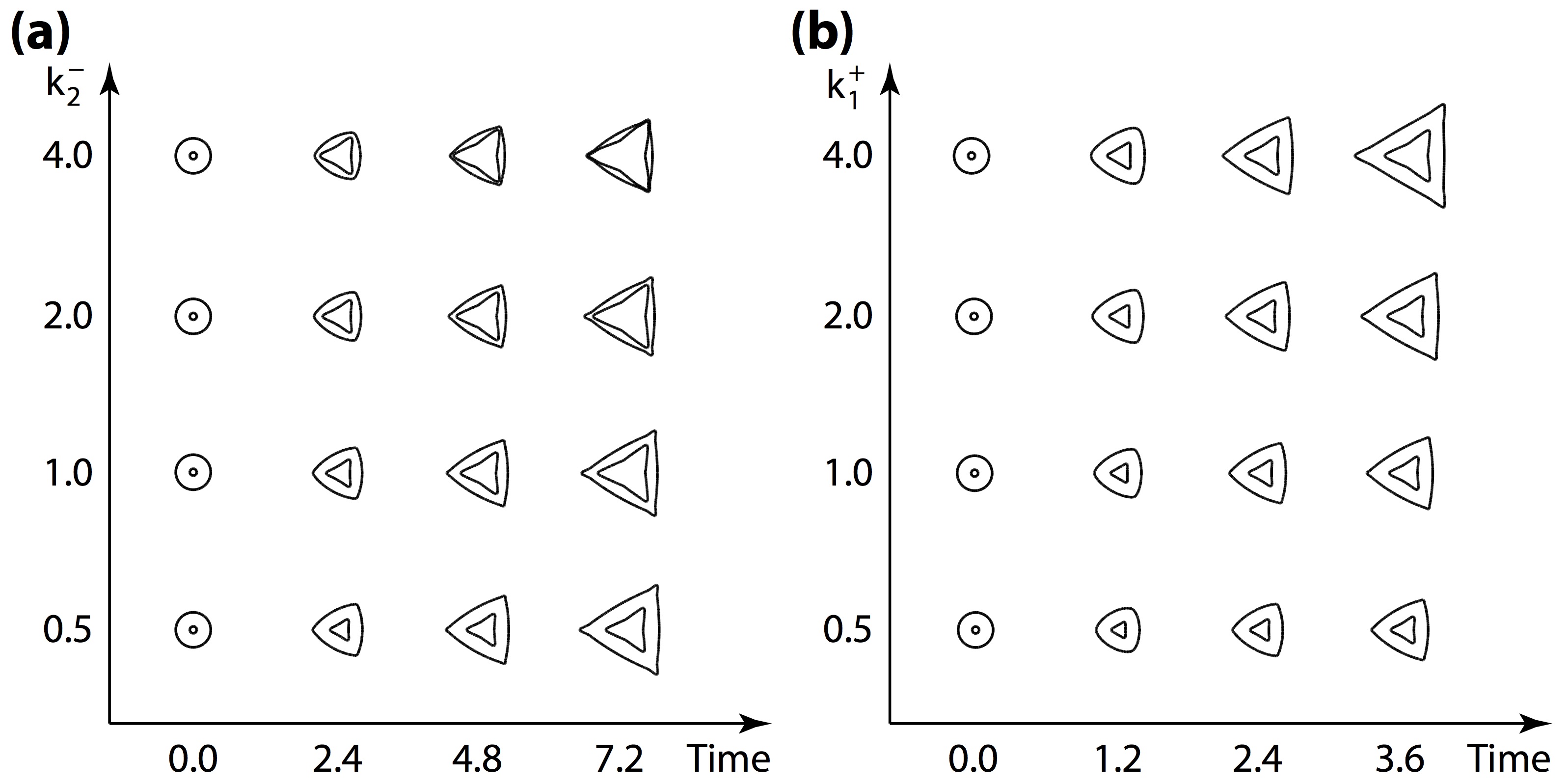}
                       \caption{The effects of kinetic attachment coefficients on the dynamics of 3-fold anisotropic layers 1 and 2. See text for parameters.  In (a), only $k^{-}_{2}$ is varied.  In (b), $k_1^+=k_1^-$ are varied together. In contrast to the 6-fold case shown in Fig. \ref{Interface-comparison-effect-k-6fold}, both kinetic parameters $k_2^-$ and $k_1^+=k_1^-$ promote the growth of layer 2 in the 3-fold case. Further, the morphological instability observed in the 6-fold case is not present in the 3-fold case.}\label{Interface-comparison-effect-k-3fold}
\end{figure}

The growth of 3-fold anisotropic layers subject to the same changes in the kinetic parameters shows somewhat different results. As seen in Figs. \ref{Interface-comparison-effect-k-3fold}(a) and (b), increasing $k_2^-$ and $k_1^+$ both favor the growth of layer 2. Further, when $k_2^-$ is increased, only layer 2 acquires negative curvature while layer 1 remains convex, in contrast to the results found for 6-fold anisotropy. In addition, when $k_1^+$ is increased, the morphological instability of layer 1 found in the 6-fold case is not present in the 3-fold case. Because of this layer 1 in the 3-fold case does not grow as rapidly, relative to that of layer 2,  which enables more adatoms to be available to drive the growth of layer 2.

\section{Conclusions}\label{Conclusion}

Epitaxial growth of 2D materials is a complex process, influenced by thermodynamic, kinetic and growth parameters, often leading to diverse and complex growth morphologies determined both by atomic-scale phenomena and by the elastic interactions of surface features and defects and transport of diffusing molecules over length scales of hundreds of nanometers. No single model can describe all the processes involved. In this paper, we derived a general continuum vdW-BCF model to describe the growth of vertically-stacked, arbitrarily-shaped multilayered 2D materials. The model accounted for (i) energy changes upon incorporation of adatoms into the growing 2D layers, (ii) kinetic barriers to attachment, (iii) distinct vdW interactions between the 2D layers and the substrate, (iv) energy penalties associated with the layer edges, and (v) the entropy of the adatoms. This is an extension of our previous work where we developed and analyzed an analogous model for faceted layers where the layer dynamics was much simpler \cite{Ye2017}. The vdW-BCF system presented here represents a highly nonlinear free boundary problem.

We analyzed a nondimensional version of the vdW-BCF model and derived an analytic thermodynamic criterion for vertical growth of stacked 2D materials assuming the layers are circular. To solve the system numerically, we used a second-order accurate phase-field/diffusion-domain method (DDM) that enabled us to solve the dynamic equations in a fixed regular domain. To discretize and solve the vdW-BCF-DDM reformulated system, we developed a second-order accurate finite-difference/nonlinear multigrid method using adaptive, block-structured Cartesian mesh refinement. We demonstrated convergence of the numerical methods and investigated the effect of parameters on the layer growth and morphological evolution. While the conditions that favor vertical growth generally follow the thermodynamic criterion we derived for circular layers, the layer boundaries may develop significant curvature during growth and even morphological instabilities. These deviations from faceted shapes can alter the growth dynamics of the layers and can hinder or enhance vertical growth.

\begin{figure}
                  \centering
                \includegraphics[width=0.9\textwidth]{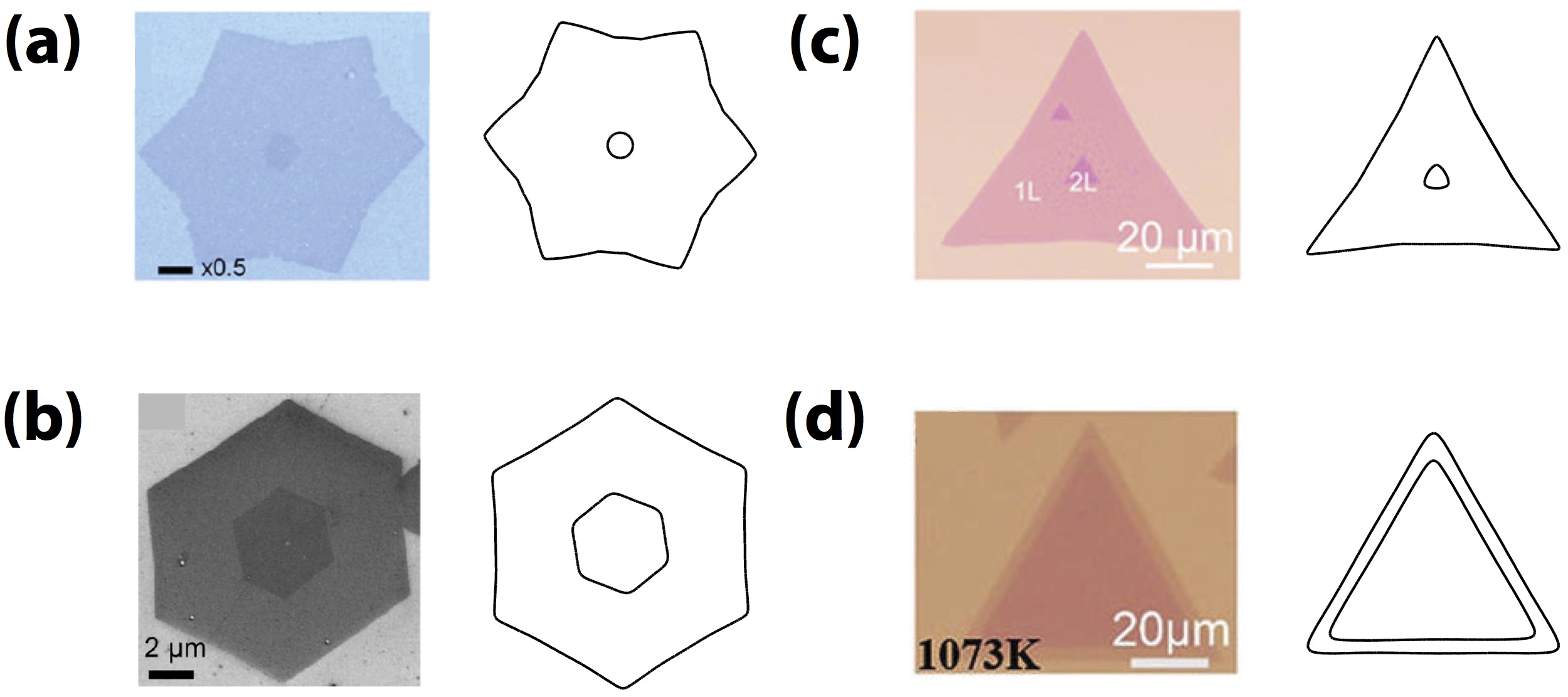}
                       \caption{Experiments show a wide variety of bilayer morphologies of 2D materials, including layers with negative curvature, which our mathematical model is capable of reproducing. (a) Left: SEM image of bilayer graphene adapted from \cite{Lu2013}, with permission. Right: Numerical simulation. (b) Left: SEM image of bilayer graphene with a twisted layer 2 adapted from \cite{Lu2013}, with permission. Right: Numerical simulation with twist angle $\tilde\theta=10^o$ (see text). (c)-(d) Left: Optical images of TMD samples showing vertically-stacked bilayers of $MoS_2$ adapted from \cite{Ye2017}, with permission. Right: Numerical simulations. See text for model parameters.}\label{numerical-compare-experiment}
\end{figure}

Experiments show a wide variety of layer morphologies, including layers with negative curvature, which our model is capable of reproducing. A small sample of experimental layer morphologies are shown in Fig.  \ref{numerical-compare-experiment} together with our numerical simulations. Fig. \ref{numerical-compare-experiment}(a) shows a SEM image of bilayer graphene from \cite{Lu2013} (left) that exhibits a star-shaped layer 1 and a nearly circular layer 2. The image on the right in Fig. \ref{numerical-compare-experiment}(a) is a numerical simulation at time $t=4$ with the parameters from Eq. (\ref{parameter-values}) except that $k^{-}_{1} = 10$, $F_{0} = 1$ and $\mathcal{E}_{2}=1.0$. Fig. \ref{numerical-compare-experiment}(b) shows a SEM image of bilayer graphene with a twisted layer 2 from \cite{Lu2013} (left). This experiment was motivated by the observation that electronic structure of bilayer graphene can be altered by changing the relative twist angle, yielding a new class of low-dimensional carbon systems. To simulate twisted bilayer graphene, we modify the reference angle $\theta_{0}$ of the kinetic coefficient $\xi_{k}(\theta) $ in Eq. (\ref{kinetic-coefficients}). In particular, we set
\begin{eqnarray}
\theta_{0}=\cases{ \frac{\pi}{6} + \frac{2\pi}{360}\times \tilde{\theta}&for $\varphi \in (1,2]$,\\
\frac{\pi}{6}&for $\varphi \in [0,1]$,\\}
\end{eqnarray}
where $\tilde{\theta}$ denotes the twist angle of layer 2. Here, we take $\tilde\theta=10^o$ and all the other parameters are as in Eq. (\ref{parameter-values}). The numerical result at time $t=8$ is shown in the right figure of Fig. \ref{numerical-compare-experiment}(b). Consistent with the experiment, layer 1 develops a hexagon shape with slight negative curvature while the twisted 2nd layer is nearly faceted. Fig. \ref{numerical-compare-experiment}(c) shows an optical image of a vertically-stacked bilayer of $MoS_2$ from \cite{Ye2017} (left) where layer 1 has a triangular shape with negatively curved sides and contains two smaller layer 2 triangles that are nearly faceted. The image on the right shows our numerical approximation at time $t=3.84$, which uses the parameters in Eq. (\ref{parameter-values}) except with $k^{\pm}_{1} =3.0$, $k^{-}_{2} = 6.0$, $\gamma_{1} =\gamma_{2} = 0.02$, and $\mathcal{E}_{2}=1.0$. Finally, in Fig. \ref{numerical-compare-experiment}(d), an optical image of a vertically-stacked bilayer of $MoS_2$ from \cite{Ye2017}(left) is shown where layer 2 nearly overlaps with layer 1 and both have shapes that are almost faceted. The figure on the right shows our numerical approximation at time $t=2.8$, which uses the parameters in Eq. (\ref{parameter-values}) except with $\gamma_{1} =\gamma_{2} = 0.04$, $F_{0} = 0.012$, and $\mathcal{E}_{2}=5.0$.

Although we performed our study using a range of nondimensional parameters, atomistic and mesoscale models can be used to provide specific material parameters. For example, DFT simulations can provide estimates for vdW interaction energies as well as edge energies and kinetic barriers for attachment \cite{Chen2015,Rajan2016,Ye2017}. Incorporating such parameter estimates will be explored in future work. 

Further, in this paper we have focused on single material homostructures due to perfect lattice matching and hence there are no interior strains. In the TMD family, one can go further and consider $MoX_2/WX_2$ heterostructures (M = Mo, W; X = S, Se, Te) without introducing lattice mismatch. However, taking full advantage of the device properties accessible through marriage of disparate 2D materials requires understanding the role of strain in the competition between vertical and in-plane lateral growth. We expect that strain-driven defect formation and stacking-site symmetry breaking will significantly modify the potential energy surface, affecting the thermodynamics of monolayer vs. multilayer morphologies and the kinetics of adatom attachment. Such effects will also be considered in future work.

\ack
One of the authors (JL) thanks Dionisios Margetis for stimulating discussions. The authors gratefully acknowledge partial support from the National Science Foundation through Grants DMS-1522775 (JL) and DMS-1522603 (VBS) as well as the Army Research Office through contract W911NF-16-1-0447.

\appendix
\section{Details of the derivation of the vdW-BCF model of vertically-stacked multilayer growth}\label{model-derivation}
\subsection{Mass Conservation}

We define the total mass to be:
\begin{equation}
M=\sum_{i=0}^2\int_{\Omega_{i}}\rho_{i}~{\rm d}A+\int_{\Omega_{1}\cup\Omega_{2}}\Omega_{s,1}~{\rm d}A+\int_{\Omega_{2}}\Omega_{s,2}~{\rm d}A,
\label{mass}
\end{equation}
where $\Omega_{s,i}$ are the concentrations of atomic sites in the layers ($i=1$, $2$). Then, mass conservation requires
\begin{equation} 
\frac{{\rm d}M}{{\rm d}t}=\sum_{i=0}^2 \left(\int_{\Omega_i} F_i~{\rm d}A- \int_{\Omega_{i}} \tau_{d,i}^{-1}\rho_i~{\rm d}A\right),
\label{mass cons 1}
\end{equation}
where $F_i$ is the deposition flux on layer $i$ and $\tau_{d,i}^{-1}$ are desorption rates. Combining these two equations and using the Reynolds transport theorem gives:
\begin{equation}
\fl \frac{{\rm d}M}{{\rm d}t}=\sum_{i=0}^2\int_{\Omega_{i}}\partial_{t} \rho_{i}+\nabla \cdot \big(\rho_{i}\mathbf{v}_i\big) {\rm d}A +\int_{\Omega_{1}\cup\Omega_{2}}\nabla\cdot\left(\Omega_{s,1}\mathbf{v}_1\right)~{\rm d}A
\int_{\Omega_{2}}\nabla\cdot\left(\Omega_{s,2}\mathbf{v}_2\right)~{\rm d}A
\label{mass cons deriv}
\end{equation}
where $\mathbf{v}_i$ are the velocities of the adatoms on the layers and substrate. For simplicity, we assume that $\Omega_{s,i}=\Omega_s$. We also assume that the boundary of the substrate $\Gamma_0$ does not move. Therefore, combining Eqs. (\ref{mass cons 1}) and (\ref{mass cons deriv}) and using the divergence theorem we obtain
\begin{equation}
0=\sum_{i=1}^2\int_{\Gamma_{i}}v_i\big(\rho_{i}^{+}  -\rho_{i}^{-} + \Omega_{s}\big)~{\rm d}A
+\sum_{i=0}^2\int_{\Omega_{i}}\left(\partial_{t} \rho_{i}-F_i+\tau_{d,i}^{-1}\rho_i\right)~{\rm d}A
\label{mass cons 2}
\end{equation}
where $v_{i}={\bf{v}}\cdot {\bf{n}}_{\Gamma_{i}}$ is the normal velocity of layer $i$, and $\rho_i^+=\rho_i|_{\Gamma_i}$, $\rho_i^-=\rho_{i-1}|_{\Gamma_i}$ are the boundary conditions for the densities at the $i$th layer from the step up and down respectively. Next, assuming that
\begin{eqnarray}
\partial_{t}\rho_{i}= -{\bf{\nabla}} \cdot {{\bf{J}}}_{i} + F_i - \tau_{d,i}^{-1}\rho_i,
\label{rho eq}
\end{eqnarray}
then the last term in Eq. (\ref{mass cons 2}) can be written as
\begin{eqnarray}
\fl\sum_{i=0}^2 \int_{\Omega_{i}}\left(\partial_{t} \rho_{i}-F_i-\tau_{d,i}^{-1}\rho_i\right)&=-\sum_{i=0}^{2}\int_{\Omega_{i}}{\bf{\nabla}} \cdot {\bf{J}}_{i} ~ {\rm d}A\nonumber\\
&=-\int_{\Gamma_{2}} (J_{2}^{+}-J_{2}^{-}){\rm d}S  -\int_{\Gamma_{1}}(J_{1}^{+}-J_{1}^{-}){\rm d}S,\label{time-variation-density}
\end{eqnarray}
where $J_{i}^{\pm}$ (for $i=1,2$) denote the fluxes at the $i$th layer from a step up and down, respectively, with
\begin{eqnarray}
{\bf{J}}_{2}\cdot {\bf{n}}_{\Gamma_{2}}&=J^{+}_{2},\hspace{10mm}{\bf{J}}_{1}\cdot {\bf{n}}_{\Gamma_{1}}&=J^{+}_{1},\\
{\bf{J}}_{1}\cdot {\bf{n}}_{\Gamma_{2}}&=J^{-}_{2}, \hspace{10mm}{\bf{J}}_{0}\cdot {\bf{n}}_{\Gamma_{1}}&=J^{-}_{1},
\end{eqnarray}
and we have assumed that there is no flux at the substrate boundary: ${\bf{J}}_{0}\cdot {\bf{n}}_{\partial\Gamma_0}=J_{0}=0$. Further, the boundary conditions for Eq. (\ref{rho eq}) on $\Gamma_i$ are taken to be
\begin{eqnarray}
q_i^+&=J_i^+-\rho_i^+v_i,
\label{bc for rho eq +}\\
q_i^-&=-J_i^-+\rho_i^-v_i.
\label{bc for rho eq -}
\end{eqnarray}
Substituting (\ref{time-variation-density}) and (\ref{bc for rho eq +})-(\ref{bc for rho eq -}),  into (\ref{mass cons 2}), we obtain
\begin{equation}
\sum_{i=1}^2\int_{\Gamma_{i}}\left(v_{i}\Omega_s-\left(q_i^++q_i^-\right)\right){\rm d}S =0.
\end{equation}
In order to satisfy mass conservation, we then have
\begin{eqnarray}
v_1&=\frac{1}{\Omega_s}\left(q_1^++q_1^--\partial_s \mathcal{J}_{1}\right),
\label{vel1}\\
v_2&=\frac{1}{\Omega_s}\left(q_2^++q_2^--\partial_s \mathcal{J}_{2}\right),
\label{vel2}
\end{eqnarray}
where $\partial_s$ denotes the arclength derivative and $\mathcal{J}_{i}$ represents surface fluxes (e.g., arising from the diffusion of adatoms along the layer edges).
To obtain constitutive laws for the fluxes $q_i$, $\mathbf{J}_i$ and $\mathcal{J}_{i}$, we require that the system dissipates the free energy when the deposition flux $F_i=0$ and desorption coefficient $\tau_{d,i}^{-1}=0$.

\subsection{Free Energy Dissipation}
Taking the time derivative of the free energy $E$ from Eq. (\ref{free energy}) and using the Reynolds transport theorem, we obtain
\begin{eqnarray}
\fl\frac{{\rm d }E}{{\rm d}t}=\sum_{i=1}^{2} \left(-\int_{\Omega_{i}}\nabla\cdot (\mathcal{E}_{i}{\bf{v}}_i)~{\rm d}A+\int_{\Gamma_{i}} {\tilde\gamma}_{i}v_{i}\kappa_{i}~{\rm d}S\right)\nonumber\\
+ \frac{k_{B}T}{\rho_{ref}}\sum_{i=0}^2 \int_{\Omega_{i}}  
\left(\partial_{t} {\rho_{i}}\left(\ln\frac{\rho_{i}}{\rho_{ref}} -\ln\left(1-\frac{\rho_{i}}{\rho_{ref}}\right)\right)\right)~{\rm d}A\nonumber\\
+k_{B}T\sum_{i=0}^2 \int_{\Omega_{i}}\nabla \cdot \bigg({\bf{v}}_i\bigg(\frac{\rho_{i}}{\rho_{ref}}\ln\frac{\rho_{i}}{\rho_{ref}}+\left(1- \frac{\rho_{i}}{\rho_{ref}}\right)\ln\left(1-\frac{\rho_{i}}{\rho_{ref}}\right)\bigg)\bigg)~{\rm d}A
\nonumber\\
\label{time deriv energy 1}
\end{eqnarray}
where $\tilde\gamma_i=\gamma_i(\theta)+\gamma^{\prime\prime}_i(\theta)$ and the primes denote derivatives with respect to $\theta$, the normal angle (e.g., angle that the normal vector makes with the $x$-axis). Defining the free energy density $f$ and the chemical potential $\mu$ to be
\begin{eqnarray}
f(\rho)&=k_BT\left(\frac{\rho}{\rho_{ref}}\ln\frac{\rho}{\rho_{ref}}+\left(1-\frac{\rho}{\rho_{ref}}\right)\ln\left(1-\frac{\rho}{\rho_{ref}}\right)\right),
\label{free energy density}\\
\mu(\rho)&=\frac{\partial f}{\partial\rho}=\frac{k_BT}{\rho_{ref}} \left(\ln\frac{\rho}{\rho_{ref}}-\ln\left(1-\frac{\rho}{\rho_{ref}}\right)\right)
\label{chemical potential}
\end{eqnarray}
and applying the divergence theorem, we obtain
\begin{eqnarray}
\fl \frac{{\rm d}E}{{\rm d}t }=\int_{\Gamma_{2}}v_{2}\big(-\mathcal{E}_{2}&+\mathcal{E}_{1}+\tilde\gamma_{2}\kappa_{2}\big){\rm d}S+\int_{\Gamma_{1}}v_{1}\big(-\mathcal{E}_{1}+\tilde\gamma_{1}\kappa_{1}\big){\rm d}S
 + \sum_{i=0}^2 \int_{\Omega_{i}}  \mu_i\partial_{t} \rho_{i} ~ {\rm d}A\nonumber\\
\fl&+\sum_{i=1}^2\int_{\Gamma_{i}}v_{i}  \left(f(\rho_i^+)-f(\rho_i^-)\right){\rm d}S,
\label{variation-energy-1}
\end{eqnarray}
where $\mu_i=\mu(\rho_i)$.
Next, using Eq. (\ref{rho eq}) in Eq. (\ref{variation-energy-1}) we obtain
\begin{eqnarray}
\frac{{\rm d}E}{{\rm d}t }&=\int_{\Gamma_{2}}v_{2}\bigg(-\mathcal{E}_{2}+\mathcal{E}_{1}+\tilde\gamma_{2}\kappa_{2}+f(\rho_2^+)-f(\rho_2^-)\bigg)
~{\rm d}S\nonumber\\
\fl &+\int_{\Gamma_{2}}v_{1}\bigg(-\mathcal{E}_{1}+\tilde\gamma_{1}\kappa_{1}+f(\rho_1^+)-f(\rho_1^-)\bigg)~
{\rm d}S\nonumber\\
\fl &\sum_{i=0}^2\left(-\int_{\Omega_{i}}\mu_{i} {\bf{\nabla}}\cdot {\bf{J}}_{i} {\rm d}A+\int_{\Omega_{i}}\mu_{i} \left(F_i-\tau_{d,i}^{-1}\rho_i\right) {\rm d}A\right).
\end{eqnarray}
Integrating by parts and using the divergence theorem, we obtain
\begin{eqnarray}
\frac{{\rm d}E}{{\rm d}t }&=\int_{\Gamma_{2}}v_{2}\bigg(-\mathcal{E}_{2}+\mathcal{E}_{1}+\tilde\gamma_{2}\kappa_{2}+f(\rho_2^+)-f(\rho_2^-)\bigg)
-(\mu_{2}^{+}J_{2}^{+}-\mu_{2}^{-}J_{2}^{-})~{\rm d}S\nonumber\\
&+\int_{\Gamma_{1}}v_{1}\bigg(-\mathcal{E}_{1}+\tilde\gamma_{1}\kappa_{1}+f(\rho_1^+)-f(\rho_1^-)\bigg)
-(\mu_{1}^{+}J_{1}^{+}-\mu_{1}^{-}J_{1}^{-})~{\rm d}S\nonumber\\
&+\sum_{i=0}^2\left(\int_{\Omega_{i}} {\bf{J}}_{i}\cdot{\bf{\nabla}}\mu_{i}~{\rm d}A+\int_{\Omega_{i}}\mu_{i} \left(F_i-\tau_{d,i}^{-1}\rho_i\right) {\rm d}A\right),
\label{energy-diss}
\end{eqnarray}
where we have defined $\mu_i^{\pm}=\mu(\rho_i^{\pm})$. See the previous subsection for the definitions of $\rho^{\pm}_i$ and $J_i^{\pm}$. Using Eqs. (\ref{bc for rho eq +}), (\ref{bc for rho eq -}),(\ref{vel1}) and (\ref{vel2}) in Eq. (\ref{energy-diss}) we obtain:
\begin{eqnarray}
\fl \frac{{\rm d}E}{{\rm d}t }&=\int_{\Gamma_{2}} q_2^+\left(\rho^{BC}_2-\mu^+_2\right)+q_2^-\left(\rho^{BC}_2-\mu^-_2\right)+\mathcal{J}_2\partial_s\rho_2^{BC}~{\rm d}S\nonumber\\
\fl &+\int_{\Gamma_{1}} q_1^+\left(\rho^{BC}_1-\mu^+_1\right)+q_1^-\left(\rho^{BC}_1-\mu^-_1\right)+\mathcal{J}_1\partial_s\rho_1^{BC}~{\rm d}S\nonumber\\
\fl & +\sum_{i=0}^2\left(\int_{\Omega_{i}} {\bf{J}}_{i}\cdot{\bf{\nabla}}\mu_{i}~{\rm d}A+\int_{\Omega_{i}}\mu_{i} \left(F_i-\tau_{d,i}^{-1}\rho_i\right) {\rm d}A\right),
\label{energy-diss-1}
\end{eqnarray}
where we have integrated by parts on the edges $\Gamma_1$ and $\Gamma_2$ and defined
\begin{eqnarray}
\rho_2^{BC}&=\frac{1}{\Omega_s}\left(-\mathcal{E}_{2}+\mathcal{E}_{1}+\tilde\gamma_{2}\kappa_{2}+L_f(\rho_2^+)-L_f(\rho_2^-)\right)
\label{rho2 BC}\\
\rho_1^{BC}&=\frac{1}{\Omega_s}\left( -\mathcal{E}_{1}+\tilde\gamma_{1}\kappa_{1}+L_f(\rho_1^+)-L_f(\rho_1^-)  \right)
\label{rho1 BC}
\end{eqnarray}
where
\begin{equation}
L_f(\rho^{\pm}_i)=f(\rho_i^{\pm})-\rho_i^{\pm}\mu_i^{\pm}=k_BT\ln\left(1-\frac{\rho_i^{\pm}}{\rho_{ref}}\right)
\label{lin approx}
\end{equation}
Hence, to have energy dissipation (in the absence of flux and desorption), we may take the constitutive relations for the fluxes:
\begin{eqnarray}
\mathbf{J}_i&=-D_i\nabla\mu_i,
\label{bulk flux}\\
\mathcal{J}_{2}&=-\beta_2\partial_s \rho_2^{BC},
\label{surface flux 2}\\
\mathcal{J}_{1}&=-\beta_1\partial_s\rho_1^{BC},
\label{surface flux 1}
\end{eqnarray}
where the $\beta_i$ are related to the mobility of an edge atom along a curved step, and the (linear) kinetic boundary conditions: 
\begin{eqnarray}
q_{2}^{\pm}&=k_{2}^{\pm}\left(\mu_2^{\pm}-\rho_2^{BC}\right),
\label{q2plus bc}\\
q_{1}^{\pm}&=k_{1}^{\pm}\left(\mu_1^{\pm}-\rho_1^{BC}\right),
\label{q1minus bc}
\end{eqnarray}
where $k_i^\pm$ are kinetic attachment coefficients.

\subsection{Model simplification}

Since $\rho_i^+ \approx \rho_i^-$ and $\rho_i^+-\rho_i^-<<\Omega_s$, we can neglect the terms $\frac{1}{\Omega_s}\left(L_f(\rho_i^+)-L_f(\rho_i^-) \right)$ in $\rho_i^{BC}$. Therefore $\rho_i^{BC}$ are approximated by
\begin{eqnarray}
\rho_2^{BC}&=\frac{1}{\Omega_s}\left(-\mathcal{E}_{2}+\mathcal{E}_{1}+\tilde\gamma_{2}\kappa_{2}\right)
\label{rho2 BC simplified}\\
\rho_1^{BC}&=\frac{1}{\Omega_s}\left( -\mathcal{E}_{1}+\tilde\gamma_{1}\kappa_{1} \right)
\label{rho1 BC simplified}
\end{eqnarray}
Further, $\mu_i^{\pm}$ can be approximated as
\begin{equation}
\mu_{i}=\frac{k_{B}T}{\rho_{ref}}\left(\ln\frac{\rho_{i}}{\rho_{ref}}-\ln\left(1-\frac{\rho_{i}}{\rho_{ref}}\right)\right)\approx \frac{4k_BT}{\rho_{ref}}\left(\frac{\rho_{i}}{\rho_{ref}}-\frac{1}{2}\right).
\label{chem pot simpl}
\end{equation}
We further neglect the effects of anisotropy in the surface fluxes (e.g., we assume that the edge energy anisotropy is small $\tilde\gamma_i(\theta)\approx \bar\gamma_i$), although we keep the effects of anisotropy in the kinetic coefficients and in $\rho_i^{BC}$.  Surface diffusion anisotropy will be considered in future work. It follows that the diffusional and surface fluxes can be approximated by
\begin{equation}
\mathbf{J}_i\approx-\tilde D_i\nabla{\rho_i}, ~~~~\mathcal{J}_{i}=-\tilde{\beta_i}\partial_s\kappa_{i},
\end{equation}
where $\tilde D_i=\frac{4D_ik_BT}{\rho_{ref}^2}$ and $\tilde\beta_i=\frac{\beta_i\bar\gamma_i}{\Omega_s}$
the velocities can be approximated as
\begin{eqnarray}
v_{1}&=\frac{1}{\Omega_s}\left(	q_{1}^{+}+q_{1}^{-}+\tilde{\beta_1}\partial_{ss}\kappa_{1}\right),
\label{simplified velocity 1 a}\\
v_{2}&=\frac{1}{\Omega_s}\left(q_{2}^{+}+q_{2}^{-}+ \tilde{\beta_2}\partial_{ss}\kappa_{2}\right),
\label{simplified velocity 2 a}
\end{eqnarray}
and the kinetic boundary conditions can be approximated as
\begin{eqnarray}
q_{2}^{+}&=\tilde k_{2}^{+}\left(\rho_2^+ - \frac{1}{\tilde\Omega_s}\left(-\tilde\mathcal{E}_{2}+\tilde\mathcal{E}_{1}+\tilde\gamma_{2}\kappa_{2}\right)\right),
\label{J2plus bc simp}\\
q_{2}^{-}&=\tilde k_{2}^{-}\left(\rho_{2}^{-}-\frac{1}{\tilde\Omega_s}\left(-\tilde\mathcal{E}_{2}+\tilde\mathcal{E}_{1}+\tilde\gamma_{2}\kappa_{2}\right)\right),
\label{J2minus bc simp}\\
q_{1}^{+}&=\tilde k_{1}^{+}\left(\rho_{1}^{+}- \frac{1}{\tilde\Omega_s}\left(-\tilde\mathcal{E}_{1}+\tilde\gamma_{1}\kappa_{1}\right)\right),
\label{J1plus bc simp}\\
q_{1}^{-} &=\tilde k_{1}^{-}\left(\rho_{1}^{-}- \frac{1}{\tilde\Omega_s}\left(-\tilde\mathcal{E}_{1}+\tilde\gamma_{1}\kappa_{1}\right)\right),
\label{J1minus bc simp}
\end{eqnarray}
where $\tilde k_i^{\pm}=k_i^\pm k_B T/\rho_{ref}^2$, ${\tilde\Omega_s}=\frac{\Omega_s k_B T}{\rho_{ref}^2}$, and $\tilde\mathcal{E}_{i}=\mathcal{E}_{i}-\frac{\rho_{ref}}{2}\tilde\Omega_s$. Finally, Eq. (\ref{rho eq}) can be approximated by
\begin{equation}
\partial_t\rho_i=\tilde D_i\Delta\rho_i+F_i-\tau_{d,i}^{-1}.
\label{nondim rho eq}
\end{equation}

\subsection{Nondimensionalization}

Let $\rho_{ref}=\Omega_s$, $\mathcal{L}$ be the characteristic size of layer $1$ and take the time scale to be $\mathcal{T}=\mathcal{L}^2\Omega^2_s/\left(4Dk_BT\right)$, where $D$ is a characteristic diffusion constant. Define the nondimensional density $\rho_i'=\rho_i/\Omega_s$ and the nondimensional flux $F_i'=\mathcal{T}F_i/\Omega_s$, where the nondimensional desorption coefficient is $\tau_{d,i}'=\tau_{d,i}/\mathcal{T}$. Then, the nondimensional adatom density equation (\ref{nondim rho eq}) becomes:
\begin{eqnarray}
\partial_{t'}\rho_{i}'= D_i'\Delta'\rho_i' + F_i' - \left(\tau_{d,i}'\right)^{-1}\rho_i',
\label{rho eq nondim}
\end{eqnarray}
where $D_i'=D_i/D$ is the nondimensional diffusion coefficient. The kinetic boundary conditions become
\begin{eqnarray}
q_{2}^{'+}&= {k}_{2}^{'+}\bigg({\rho}_{2}' - {\rho}^*(-\mathcal{E}_{2}'+\mathcal{E}_{1}'+\tilde\gamma_{2}'\kappa_2')\bigg),\\
q_{2}^{'-}&=- {k}_{2}^{'-}\bigg({\rho}_{1}'-{\rho}^*(-\mathcal{E}_{2}'+\mathcal{E}_{1}'+\tilde\gamma_{2}'\kappa_{2}')\bigg),\\
q_{1}^{'+}&={k}_{1}^{'+}\bigg({{\rho}}_{1}'-{ \rho}^*(-\mathcal{E}_{1}'+\tilde\gamma_{1}'\kappa_{1}')\bigg),\\
q_{1}^{'-}&=-{ k}_{1}^{'-}\bigg({\rho}_{0}'-{\rho}^*(-\mathcal{E}_{1}'+\tilde\gamma_{1}'\kappa_{1}')\bigg).
\end{eqnarray}
where 
\begin{eqnarray}
\fl q^{'\pm}_i=\frac{\mathcal{L}\rho_{ref}}{4Dk_BT}q^{\pm}_i,~~k^{'\pm}_{i}=k^{\pm}_{1,2}\frac{\mathcal{L}\rho_{ref}}{D},~~{\rho}^{*}=\frac{\mathcal{E}}{4k_{B}T},~~\mathcal{E}_i'=\mathcal{E}_i/\mathcal{E}-\frac{1}{2\rho^*},~~\tilde\gamma_i'=\tilde\gamma_i/\left(\mathcal{E}\mathcal{L}\right),
\nonumber\\
\end{eqnarray}
and $\mathcal{E}$ is a characteristic value of the binding energies.
Finally, the nondimensional velocities are:
\begin{eqnarray}
v_{1}'&=q_{1}^{'+}+q_{1}^{'-}+\beta_1'\partial_{s's'}\kappa_{1}',
\label{simplified velocity 1}\\
v_{2}'&=q_{2}^{'+}+q_{2}^{'-}+ \beta_2'\partial_{s's'}\kappa_{2}',
\label{simplified velocity 2 nondim}
\end{eqnarray}
where $\beta_i'=\frac{\beta_i\bar\gamma_i}{4Dk_BT\mathcal{L}^2}$
are nondimensional edge diffusion coefficients. Dropping the primes, this is the system given in Sec. \ref{model equations}.

\subsection{The vdW-BCF model equations for an arbitrary number of vertically-stacked layers}
\label{arbitrary layers}

One can extend the vdW-BCF model derived in the previous sections to describe the dynamics of an arbitrary number of layers. The resulting (nondimensional) system is 
\begin{eqnarray}
\partial_{t} \rho_{i}= D_{i}\Delta \rho_{i} +F_{i} -\tau^{-1}_{d} \rho_{i}   ~~~~{\rm in}~~\Omega_{i}, ~~~~i=0,1,\dots n,\label{bcf-multiple-layer}
\end{eqnarray}
where $n$ is the number of layers.
The boundary conditions at the boundary of the first layer with the substrate, $\Gamma_{1}$, are given as
\begin{eqnarray}
q_{1}^{+}&=-D_{1}{\bf{\nabla}}\rho_{1}\cdot {\bf{n}}_{\Gamma_{1}}-\rho_1|_{\Gamma_1}  =k_{1}^{+}\bigg({\rho}_{1}- \rho^*(-\mathcal{E}_{1}+\gamma_{1}\kappa_{1})\bigg),\label{Multi-bc31} \\
q_{1}^{-}&=\hspace{3mm}D_{0}{\bf{\nabla}}\rho_{0}\cdot {\bf{n}}_{\Gamma_{1}}+\rho_0|_{\Gamma_1}= k_{1}^{-}\bigg(\rho_{0}-\rho^*(-\mathcal{E}_{1}+\gamma_{1}\kappa_{1})\bigg),\label{Multi-bc41} 
\end{eqnarray}
and for all the layer boundaries (e.g., steps) $\Gamma_{i}$ (for $i=2,\dots, n$) are:
\begin{eqnarray}
q_{i}^{+}&=\hspace{7.5mm}-D_{i}{\bf{\nabla}}\rho_{i}\cdot {\bf{n}}_{\Gamma_{i}}-\rho_i|_{\Gamma_i}= k_{i}^{+}\bigg({\rho}_{i} - \rho^*(-\mathcal{E}_{i}+\mathcal{E}_{i-1}+\gamma_{i}\kappa_i)\bigg),\label{Multi-bc11} \\
 q_{i}^{-}&=\hspace{3mm}D_{i-1}{\bf{\nabla}}\rho_{i-1}\cdot {\bf{n}}_{\Gamma_{i}}+\rho_{i-1}|_{\Gamma_i}  = k_{i}^{-}\bigg(\rho_{i-1}-\rho^*(-\mathcal{E}_{i}+\mathcal{E}_{i-1}+\gamma_{i}\kappa_{i})\bigg),
\nonumber \\
 \label{Multi-bc21} 
\end{eqnarray}
where $\gamma_{i}$ denotes the step stiffness and $\kappa_{i}$ is the curvature of the $i$th step $\Gamma_{i}$, for $i=1,2\dots n$. The normal velocity of each step $\Gamma_{i}$ is given by
\begin{eqnarray}
v_{i}=q_{i}^{+}+q_{i}^{-}+\beta \partial_{s}^{2}\kappa_{i}.\label{surface-diffusion-multi}
\end{eqnarray}

\section{Details of the derivation of radial solutions to the vdW-BCF model}\label{model-analysis details}
We now derive the analytic solutions $\rho_{i}(r,t)$ in the quasi-steady state limit. That is, we drop the time derivatives in the adatom diffusion equations. We first rewrite Eq. (\ref{bcf-double-layer}) as
\begin{eqnarray}
-\frac{D_{i}}{r}\partial_{r}(r\partial_{r}\rho_{i}) = F_{i},
\end{eqnarray}
where we have also neglected desorption and taken $\tau_{d,i}^{-1}=0$.
Integrating twice we obtain:
\begin{eqnarray}
\rho _{i}= - \frac{F_{i}}{4D_{i}}r^{2} + A_{i}{\rm ln}(r) +B_{i},~~~~{\rm for }~i=0,1,2,\label{BCF-ana-exp}
\end{eqnarray}
where $A_{i}$ and $B_{i}$ are unknown constants.\\
For $r<R_{2}(t)$, the solution in Eq. (\ref{BCF-ana-exp}) satisfies the following boundary conditions:
\begin{eqnarray}
\rho_{2}~{\rm is~continuous},~~~~&{\rm at}~r = 0,\label{BCF-BC-1-1}\\   
-D_{2}\partial_{r}\rho_{2}= k^{+}_{2}\big[\rho_{2}-\rho^*(-\mathcal{E}_{2}+\mathcal{E}_{1}+\frac{\gamma_{2}}{R_{2}} )\big],~~~~&{\rm at}~r = R_{2}.\label{BCF-BC-1-2}
\end{eqnarray}
We then obtain 
\begin{eqnarray}
A_{2}&=0,~~~~B_{2} = \frac{F_{2}R_{2}^{2}}{4D_{2}}  +\frac{F_{2}R_{2}}{2k^{+}_{2}}  + \rho^*(-\mathcal{E}_{2}+\mathcal{E}_{1}+\frac{\gamma_{2}}{R_{2}}).
\end{eqnarray}
For $R_{2}<r<R_{1}$, the solution in Eq. (\ref{BCF-ana-exp}) satisfies 
\begin{eqnarray}
D_{1}\partial_{r}\rho_{1} = k^{-}_{2}\big[\rho_{1}-\rho^*(-\mathcal{E}_{2}+\mathcal{E}_{1}+\frac{\gamma_{2}}{R_{2}})\big],~~~~&{\rm at}~r = R_{2},\label{2BCF-BC-2-1}\\
-D_{1}\partial_{r}\rho_{1} = k^{+}_{1}\big[\rho_{1}-\rho^*(-\mathcal{E}_{1}+\frac{\gamma_{1}}{R_{1}} )\big],~~~~&{\rm at}~r = R_{1}.\label{2BCF-BC-2-2}
\end{eqnarray}
At $r = R_2$, we obtain
\begin{eqnarray}
&B_{1}=\frac{F_{1}R_2^{2}}{4D_{1}} - A_{1}\ln{R_2} + \frac{D_{1}}{k^{-}_{2}}(-\frac{F_{1}R_2}{2D_{1}}+\frac{A_{1}}{R_2})+ \rho^*(-\mathcal{E}_{2}+\mathcal{E}_{1}+\frac{\gamma_{2}}{R_2}).
\end{eqnarray}
At $r = R_1$, we obtain
\begin{eqnarray}
\fl B_{1}=\frac{F_{1}R_1^{2}}{4D_{1}} - A_{1}\ln{R_1}  -\frac{D_{1}}{k^{+}_{1}}(-\frac{F_{1}R_1}{2D_{1}}+\frac{A_{1}}{R_1}) +\rho^*(- \mathcal{E}_{1}+\frac{\gamma_{1}}{R_1}),
\end{eqnarray}
such that 
\begin{eqnarray}
\fl A_{1}=\frac{\frac{F_{1}}{4D_{1}}(R_2^{2}-R_1^{2}) + \rho^*(-\mathcal{E}_{2}+\mathcal{E}_{1}+\frac{\gamma_{2}}{R_2}) -\frac{F_{1}R_1}{2k^{+}_{1}} -\frac{F_{1}R_2}{2k^{-}_{2}}-\rho^*(-\mathcal{E}_{1}+\frac{\gamma_{1}}{R_1})}{\bigg(\ln{\frac{R_2}{R_1}}-\frac{D_{1}}{k^{-}_{2}R_2}- \frac{D_{1}}{k^{+}_{1}R_{1}}\bigg)}.
\end{eqnarray}
For $R_{1}<r<R_\infty$, the solution in Eq. (\ref{BCF-ana-exp}) satisfies 
\begin{eqnarray}
D_{0}\partial_{r}\rho_{0} = k^{-}_{1}\bigg(\rho_{0}-\rho^*(-\mathcal{E}_{1}+\frac{\gamma_{1}}{R_{1}})\bigg),~~~~&{\rm at}~r = R_1,\label{BCF-BC-2-1}\\
\partial_{r}\rho_{0} = 0,~~~~&{\rm at}~r = R_\infty.\label{BCF-BC-2-2}
\end{eqnarray}
At $r = R_1$, we obtain
\begin{eqnarray}
-\frac{F_{0}R_{1}^2}{4D_{0}} + A_{0}\ln{R_{1}} +B_{0} -\frac{D_{0}}{k^{-}_{1}}(-\frac{F_{0}R_{1}}{2D_{0}}+\frac{A_{0}}{R_{1}}) &=\rho^*(-\mathcal{E}_{1}+\frac{\gamma_{1}}{R_{1}})
\end{eqnarray}
At $r = R_{0}$, we obtain
\begin{eqnarray}
A_{0}= \frac{F_{0}}{2D_{0}}R^2_{0}.
\end{eqnarray}
such that
\begin{eqnarray}
B_{0}=\frac{F_{0}R_{1}^2}{4D_{0}} - \frac{F_{0}R^2_{0}}{2D_{0}}\ln{R_{1}} - \frac{F_{0}R_{1}}{2k^{-}_{1}} +\frac{F_{0}R^2_{0}}{2k^{-}_{0}R_{1}} +\rho^*(-\mathcal{E}_{1}+\frac{\gamma_{1}}{R_{1}}).
\end{eqnarray}
Summarizing, we obtain the analytic solution
\begin{eqnarray}
{\rho} _{2}&= - \frac{F_{2}}{4D_{2}}r^{2} + A_{2}{\rm ln}(r) +B_{2},\hspace{14.5mm}x<R_{2},\nonumber\\
{\rho} _{1}&= - \frac{F_{1}}{4D_{1}}r^{2} + A_{1}{\rm ln}(r) +B_{1},\hspace{5mm}R_{2}<x<R_{1},\nonumber\\
{\rho} _{0}&= - \frac{F_{0}}{4D_{0}}r^{2} + A_{0}{\rm ln}(r) +B_{0},\hspace{5mm}R_{1}<x,\label{analytical-solution}
\end{eqnarray}
where
\begin{eqnarray}
A_{2}&=0,\nonumber\\
B_{2} &= \frac{F_{2}R_{2}^{2}}{4D_{2}}  +\frac{F_{2}R_{2}}{2k^{+}_{2}}  + \rho^*(-\mathcal{E}_{2}+\mathcal{E}_{1}+\frac{\gamma_{2}}{R_{2}}),\nonumber\\
A_{1}&=\frac{\frac{F_{1}}{4D_{1}}(R_2^{2}-R_1^{2}) + \rho^*(-\mathcal{E}_{2}+\mathcal{E}_{1}+\frac{\gamma_{2}}{R_2}) -\frac{F_{1}R_1}{2k^{+}_{1}} -\frac{F_{1}R_2}{2k^{-}_{2}}-\rho^*(-\mathcal{E}_{1}+\frac{\gamma_{1}}{R_1})}{\bigg(\ln{\frac{R_2}{R_1}}-\frac{D_{1}}{k^{-}_{2}R_2}- \frac{D_{1}}{k^{+}_{1}R_{1}}\bigg)},\nonumber\\
B_{1}&=\frac{F_{1}R_1^{2}}{4D_{1}} - A_{1}\ln{R_1}  -\frac{D_{1}}{k^{+}_{1}}(-\frac{F_{1}R_1}{2D_{1}}+\frac{A_{1}}{R_1}) +\rho^*(- \mathcal{E}_{1}+\frac{\gamma_{1}}{R_1}),\nonumber\\
A_{0}&= \frac{F_{0}}{2D_{0}}R^2_{0},\nonumber\\
B_{0}&=\frac{F_{0}R_{1}^2}{4D_{0}} - \frac{F_{0}R^2_{0}}{2D_{0}}\ln{R_{1}} - \frac{F_{0}R_{1}}{2k^{-}_{1}} +\frac{F_{0}R^2_{0}}{2k^{-}_{0}R_{1}} +\rho^*(-\mathcal{E}_{1}+\frac{\gamma_{1}}{R_{1}}).\nonumber
\end{eqnarray}
The corresponding velocities of the layer boundaries are
\begin{eqnarray}
v_2&=-\left(D_2\partial_r\rho_2-D_1\partial_r\rho_1\right)|_{r=R_2}=\frac{R_2\left(F_2-F_1\right)}{2}+\frac{D_1A_1}{R_2},
\label{layer velocity 1 appendix}\\
v_1&=-\left(D_1\partial_r\rho_1-D_0\partial_r\rho_0\right)|_{r=R_1}=\frac{R_1\left(F_1-F_0\right)}{2}+\frac{D_0A_0-D_1A_1}{R_1}.
\end{eqnarray}

\section{The diffuse domain method: Details and asymptotic analysis}
\label{diffuse domain method details}

For simplicity, consider the problem with a single layer:
\begin{eqnarray}
\partial_t\rho_i=D_i\Delta\rho_i + F_i - \tau^{-1}\rho_i,~~{\rm in}~~\Omega_i(t)
\label{diff 1}
\end{eqnarray}
where $i=0$, $1$ denote the substrate and layer, respectively. The kinetic boundary conditions are:
\begin{eqnarray}
q_1^+&=-D_1\nabla\rho_1\cdot\mathbf{n}_1-\rho_1v_1=k_1^+\left(\rho_1-\rho^*\left(-\mathcal{E}_1+\tilde\gamma_1\kappa\right)\right)
\label{sbc 1}\\
q_1^-&=D_0\nabla\rho_0\cdot\mathbf{n}_1+\rho_0 v_1=k_1^-\left(\rho_0-\rho^*\left(-\mathcal{E}_1+\tilde\gamma_1\kappa\right)\right)
\label{sbc 2}
\end{eqnarray}
with the normal velocity of $\Gamma_1(t)=\partial\Omega_1(t)$ given by
\begin{eqnarray}
v_1=q_1^++q_1^-+\beta\partial_s^2\kappa.
\label{s normal vel}
\end{eqnarray}
In the above, $\kappa$ is the curvature of $\Gamma_1$.

Next, following  \cite{Diffuse-Domain,000356209300006}, we can reformulate Eqs. (\ref{diff 1})-(\ref{sbc 2}) as
\begin{eqnarray}
&\partial_t\left(\varphi\rho^\epsilon_1\right)=\nabla\cdot\left(D_1\varphi\nabla\rho^\epsilon_1\right)+\varphi \left(F_1-\tau^{-1}\rho^\epsilon_1\right)
 -k_1^+|\nabla\varphi|\left(\rho^\epsilon_1-g\right),
\label{d diff 1}\\
&\partial_t\left(\varphi^c\rho^\epsilon_0\right)=\nabla\cdot\left(D_0\varphi^c\nabla\rho^\epsilon_0\right)+\varphi^c \left(F_0-\tau^{-1}\rho^\epsilon_0\right) -k_1^-|\nabla\varphi|\left(\rho^\epsilon_0-g\right),
\label{d diff 2}\\
&g=\rho^*\left(-\mathcal{E}_1+\epsilon^{-1}\tilde\gamma_1\mu\right),
\label{d diff 3}
\end{eqnarray}
where $\varphi=\varphi(\mathbf{x},t)$ is a phase-field function that approximates the characteristic function of $\Omega_1(t)$, $\varphi^c=1-\varphi$ approximates the characteristic function of the substrate $\Omega_0$, and $\mu=B'(\varphi)-\epsilon^2\Delta\varphi$ is the chemical potential where $B(\varphi)=18\varphi^2\left(1-\varphi\right)^2$ is a double well free energy. Eqs. (\ref{d diff 1}) and (\ref{d diff 2}) are solved in a large rectangular domain $\tilde\Omega$ that contains $\Omega_1$ and $\Omega_2$. For simplicity, we do not include $\varphi_\infty$ to specify that the deposition domain on the substrate is a circle and we assume that the kinetic parameters and edge energies are isotropic. The evolution of the layer is captured by the Cahn-Hilliard-like model:
\begin{eqnarray}
\partial_t\varphi&=|\nabla\varphi|\left(k_1^+\left(\rho^\epsilon_1-g\right)+k_1^- \left(\rho^\epsilon_0-g\right)\right)+\frac{\beta}{\epsilon^2}\nabla\cdot\left(G(\phi)\nabla\mu\right),
\label{ch 1}\\
\mu&= B'(\varphi)-\epsilon^2\Delta\varphi,
\label{ch 2}\\
B(\varphi)&=18\varphi^2\left(1-\varphi\right)^2
\label{ch 3}\\
G(\varphi)&=2 B(\varphi).
\label{ch 4}
\end{eqnarray}

Below, we demonstrate using the method of matched asymptotic expansions that the DDM (\ref{d diff 1})-(\ref{ch 4}) yields a second-order accurate approximation of the sharp interface system (\ref{diff 1})-(\ref{s normal vel}). The analysis can easily be extended to the more complete model presented in the main text in Sec. \ref{DDM} where two layers are considered and the substrate geometry is circular (implemented via $\varphi_\infty$).

\paragraph{Matched asymptotic expansions.}
Away from the layer 1 boundary $\Gamma_1(t)$, we assume that all variables are smooth and have regular expansions in $\epsilon$, e.g., 
\begin{equation}
\rho^\epsilon_i=\rho^{(0)}_i+\epsilon\rho^{(1)}_i+\epsilon^2\rho^{(2)}_i+\dots,
\label{regular expansion}
\end{equation}
while away from $\Gamma_1$, $\varphi= 1$ inside $\Omega_1$ and $\varphi=0$ outside $\Omega_1$ to all orders. Accordingly, we see that $\rho^{(0)}_i$ satisfies Eq. (\ref{diff 1}), while the first order perturbations satisfy:
\begin{eqnarray}
\partial_t\rho^{(1)}_i=D_i\Delta\rho^{(1)}_i - \tau^{-1}\rho^{(1)}_i~~{\rm in}~~\Omega_i(t).
\label{diff 1 pert}
\end{eqnarray}
To provide the boundary conditions for the diffusion equations, we need to analyze the behavior of the system near $\Gamma_i$. To argue that $\rho^\epsilon_i$ is a second order approximation to the sharp interface solution $\rho_i$, we need to demonstrate that $\rho^{(1)}_i=0$. 

Near $\Gamma_i$, we introduce a  stretched, local coordinate system:
\begin{equation}
\mathbf{x}({s},t;\epsilon)=\mathbf{X}({s},t)+\epsilon z\mathbf{n}({s},t)
\label{local coordinate system}
\end{equation}
where $\mathbf{X}({s},t)$ is a parameterization of $\Gamma_1(t)$, $s$ is arclength, $\mathbf{n}(\mathbf{x},t)$ is the normal vector that points out of $\Omega_1$, $z=r(\mathbf{x},t)/\epsilon$ is a stretched normal coordinate and $r(\mathbf{x},t)$ is the signed distance from $\mathbf{x}$ to $\Gamma_1(t)$. In the local coordinate system, derivatives become:
\begin{eqnarray}
\mathbf{\nabla}&=\frac{1}{\epsilon}\mathbf{n}\partial_z+\mathbf{s}\frac{1}{1+\epsilon z\kappa}\partial_s
\label{grad}\\
\Delta &=\frac{1}{\epsilon^2}\partial_{zz}+\frac{1}{\epsilon}\frac{\kappa}{1+\epsilon z\kappa}\partial_z+\frac{1}{1+\epsilon z\kappa}\partial_s\left(\frac{1}{1+\epsilon z\kappa}\partial_s\right)
\label{laplacian}\\
\partial_t &= -\frac{v^\epsilon_1}{\epsilon}\partial_z+\partial_t,
\label{time deriv}
\end{eqnarray}
where the time derivative on the left hand side of Eq. (\ref{time deriv}) is the full time derivative and the time derivative on the right hand side is the time partial derivative in the inner variables, and $v_1^\epsilon$ is the effective diffuse interface normal velocity of $\Gamma_1$.
Note that $\mathbf{n}=-\nabla\varphi/|\nabla\varphi|$. We assume that near $\Gamma_1(t)$, the inner variables can be expressed as,
\begin{equation}
\hat\rho^{\epsilon}_i(z,s,t)=\rho^{\epsilon}_i(\mathbf{X}({s},t)+\epsilon z\mathbf{n}({s},t),t).
\label{local system}
\end{equation}
We assume that in the inner expansion, all variables have a regular expansion in the stretched coordinates, e.g.,
\begin{equation}
\hat\rho^{\epsilon}_i(z,s,t)=\hat\rho^{(0)}_i(z,s,t)+\epsilon \hat\rho^{(1)}_i(z,s,t)+\epsilon^2\hat\rho^{(2)}_i(z,s,t)+\dots
\label{inner expansion}
\end{equation}
To match the inner and outer expansions, we assume that there is a region of overlap where both expansions are valid and must match. In particular, if we evaluate the outer solution in the inner variables, this must match the limits of the inner solutions away from the interface. That is,
\begin{equation}
\rho^{\epsilon}_i(\mathbf{X}+\epsilon z\mathbf{n},t)\sim \hat\rho^{\epsilon}_i(z,s,t),
\label{matching 1}
\end{equation}
as $z\to\pm\infty$ and $\epsilon\to 0$ with $\epsilon z\to 0^\pm$.
Using the inner and outer expansions and equating the powers of $\epsilon$, we obtain
\begin{eqnarray}
\hat\rho^{(0)}_i(z,s,t) &\sim \rho^{(0)}_i(s,t),
\label{match 1}\\
\hat\rho^{(1)}_i(z,s,t) &\sim \rho^{(1)}_i(s,t)+z\mathbf{n}\cdot\nabla \rho^{(0)}_i(s,t),
\label{match 2}\\
\hat\rho^{(2)}_i(z,s,t) &\sim \rho^{(2)}_i(s,t)+z\mathbf{n}\cdot\nabla \rho^{(1)}_i(s,t)+\frac{z^2}{2}\mathbf{n}\cdot\nabla\nabla\rho^{(0)}_i(s,t)\cdot\mathbf{n},
\label{match 3}\\
&\vdots\nonumber
\end{eqnarray}
where $\rho^{(k)}_i(s,t)=\rho^{(k)}_i(\mathbf{X}(s,t),t)$.

Next, transforming the equations, plugging in the inner expansions and equating powers of $\epsilon$ we derive equations governing the inner solutions. At leading order $O(\epsilon^{-2})$, we obtain
\begin{eqnarray}
& \partial_z\left(\hat\varphi^{(0)}\partial_z\hat\rho^{(0)}_1\right)=0,
\label{inner 1a}\\
& \partial_z\left((1-\hat\varphi^0)\partial_z\hat\rho^{(0)}_0\right)=0
\label{inner 1b}
\end{eqnarray}
From these equations (and the matching conditions), we conclude that 
\begin{equation}
\partial_z\hat\rho^{(0)}_1=\partial_z\hat\rho^{(0)}_0=0,
\label{inner 1c}
\end{equation}
so that $\hat\rho^{(0)}_0$ and $\hat\rho^{(0)}_1$ are constant in $z$ across the inner layer. At the next order $O(\epsilon^{-1})$ we obtain:
\begin{eqnarray}
&-v_1^{(0)}\partial_z\left(\hat\varphi^{(0)}\hat\rho^{(0)}_1\right)=D_1\partial_z\left(\hat\varphi^{(0)}\partial_z\hat\rho^{(1)}_1\right)+k_1^+\left(\hat\rho^{(0)}_1-\hat g^{(0)}\right)\partial_z\hat\varphi^{(0)}
\label{inner 2a}\\
&-v_1^{(0)}\partial_z\left((1-\hat\varphi^{(0)})\hat\rho^{(0)}_0\right)=D_0\partial_z\left((1-\hat\varphi^{(0)})\partial_z\hat\rho^{(1)}_0\right)+k_1^-\left(\hat\rho^{(0)}_0-\hat g^{(0)}\right)\partial_z\hat\varphi^{(0)}
\nonumber\\
\label{inner 2b}
\end{eqnarray}
Integrating these equations from $-\infty$ to $\infty$ in $z$, using that $v_1^{(0)}$ is independent of $z$, $\hat\varphi^{(0)}(+\infty)=0$ and  $\hat\varphi^{(0)}(+\infty)=1$, we obtain
\begin{eqnarray}
&D_1\partial_z\hat\rho^{(1)}_1(-\infty)+v_1^{(0)}\hat\rho^{(0)}_1=-k_1^+\left(\hat\rho^{(0)}_1-\hat g^{(0)}\right),
\label{inner 2c}\\
-&D_0\partial_z\hat\rho^{(1)}_0(-\infty)-v_1^{(0)}\hat\rho^{(0)}_0=-k_1^0\left(\hat\rho^{(0)}_0-\hat g^{(0)}\right),
\label{inner 2d}
\end{eqnarray}
where we have additionally used Eq. (\ref{inner 1c}) and assumed that $\partial_z\hat g^{(0)}=0$, a fact that will be justified later. From the matching conditions Eqs. (\ref{match 1}) and (\ref{match 2}), we obtain
\begin{eqnarray}
-&D_1\mathbf{n}\cdot\nabla\rho^{(0)}_1-v_1^{(0)}\rho^{(0)}_1=k_1^+\left(\rho^{(0)}_1- g^{(0)}\right),
\label{inner 2cc}\\
&D_0\mathbf{n}\cdot\nabla\rho^{(0)}_0+v_1^{(0)}\rho^{(0)}_0=k_1^0\left(\rho^{(0)}_0-g^{(0)}\right),
\label{inner 2dd}
\end{eqnarray}
where, as stated earlier, $\rho^{(0)}_i$ are the limiting values of the leading order outer solution on $\Omega_i$ and we have defined $g^{(0)}=\hat g^{(0)}$. Now, using that $g^{(0)}=\rho^*\left(-\mathcal{E}_1-\tilde\gamma_1\kappa\right)$, another fact we will demonstrate later, then we recover the kinetic boundary conditions Eqs. (\ref{sbc 1}) and (\ref{sbc 2}). This implies that $\rho^{(0)}_i$ satisfies the sharp interface diffusion equations and kinetic boundary conditions, e.g., Eqs. (\ref{diff 1})-(\ref{sbc 2}). 

To justify the assumptions for $\hat g^{(0)}$ and to determine the normal velocity $v_1^{(0)}$, we need to analyze the Cahn-Hilliard-like system (\ref{ch 1}) and (\ref{ch 2}). Before doing this, however, we proceed to the next order in the inner expansion for the adatom diffusion equations in order to determine the boundary conditions for Eq. (\ref{diff 1 pert}) for the outer solution at the next order, $\rho^{(1)}_i$. At $O(1)$, and after manipulation, we obtain
\begin{eqnarray}
-&\partial_z\left(v_1^{(1)}\hat\varphi^{(0)}\hat\rho^{(0)}_1+v_1^{(0)}\hat\varphi^{(0)}\hat\rho^{(1)}_1\right)+\partial_t\left(\varphi^{(0)}\hat\rho^{(0)}_1\right)\nonumber\\
&=D_1\left(\partial_z\left(\hat\varphi^{(0)}\partial_z\hat\rho^{(2)}_1\right)+\kappa\hat\varphi^{(0)}\partial_z\hat\rho^{(1)}_1+\hat\varphi^{(0)}\partial_{ss}\hat\rho^{(0)}_1\right)
\nonumber\\
&\qquad\qquad\qquad\qquad\qquad +k_1^+\left(\hat\rho^{(1)}_1-\hat g^{(1)} \right) \partial_z\hat\varphi^{(0)}+\hat\varphi^{(0)}F_1,
\label{inner 3a}
\end{eqnarray}
and
\begin{eqnarray}
&-\partial_z\left(v_1^{(1)}\left(1-\hat\varphi^{(0)}\right)\hat\rho^{(0)}_0+v_1^{(0)}\left(1-\hat\varphi^{(0)}\right)\hat\rho^{(1)}_0\right)+\partial_t\left(\left(1-\varphi^{(0)}\right)\hat\rho^{(0}_0\right)\nonumber\\
&=D_0\left(\partial_z\left(\left(1-\hat\varphi^{(0)}\right)\partial_z\hat\rho^{(2)}_0\right)+\kappa\left(1-\hat\varphi^{(0)}\right)\partial_z\hat\rho^{(1)}_0+\left(1-\hat\varphi^{(0)}\right)\partial_{ss}\hat\rho^{(0)}_0\right)
\nonumber\\
&\qquad\qquad\qquad\qquad\qquad +k_1^-\left(\hat\rho^{(1)}_0-\hat g^{(1)} \right) \partial_z\hat\varphi^{(0)}+\left(1-\hat\varphi^{(0)}\right)F_0,
\label{inner 3b}
\end{eqnarray}
where we have used
\begin{eqnarray}
&\partial_z\left(\left(-v^{(0)}\hat\rho^{(0)}_1-D_1\partial_z\hat\rho^{(1)}_1-k_1^+\left(\hat\rho^{(0)}_1-\hat g^{(0)}\right)\right)\hat\varphi^{(0)}\right)=0,
\label{inner 3c}\\
&\partial_z\left(\left(-v^{(0)}\hat\rho^{(0)}_0-D_0\partial_z\hat\rho^{(1)}_0-k_1^-\left(\hat\rho^{(0)}_0-\hat g^{(0)}\right)\right)\left(1-\hat\varphi^{(0)}\right)\right)=0,
\label{inner 3d}
\end{eqnarray}
which follow from Eqs. (\ref{inner 2a}) and (\ref{inner 2b}) and using the matching conditions. Next, we observe that on $\Gamma_i$:
\begin{eqnarray}
-D_i\mathbf{n}\cdot\nabla\nabla\rho^{(0)}_i\cdot\mathbf{n}&=D_i\left(\Delta\rho^{(0)}_i-\kappa\mathbf{n}\cdot\nabla\rho^{(0)}_i-\partial_{ss}\rho^{(0)}_i\right)
\label{inner 4a}\\
&=\partial_t\rho^{(0)}_i-F_i-D_i\left(\kappa\mathbf{n}\cdot\nabla\rho^{(0)}_i+\partial_{ss}\rho^{(0)}_i\right),
\label{inner 4b}
\end{eqnarray}
where we have used that $\rho^{(0)}_i$ satisfies Eq. (\ref{diff 1}). Using this in the matching conditions (\ref{match 1})-(\ref{match 3}), we obtain:
\begin{eqnarray}
&-D_1\mathbf{n}\cdot\nabla \rho^{(1)}_1-k_1^+ \rho^{(1)}_1 -v^{(0)}\rho^{(1)}_1\sim -D_1\partial_z\hat\rho^{(2)}_1-k_1^+\hat\rho^{(1)}_1-v^{(0)}\hat\rho^{(1)}_1
\nonumber\\
&\qquad\qquad+\left(k_1^++v^{(0)}\right)z\mathbf{n}\cdot\nabla\rho^{(0)}_1+z\left(\partial_t\rho^{(0)}_1-F_1-D_1\left(\kappa\mathbf{n}\cdot\nabla\rho^{(0)}_1+\partial_{ss}\rho^{(0)}_1\right)\right).
\nonumber\\
\label{inner 5}
\end{eqnarray}
This motivates us to rewrite Eq. (\ref{inner 3a}) as
\begin{eqnarray}
-&\partial_z\left(D_1\hat\varphi^{(0)}\partial_z\hat\rho^{(2)}_1+v_1^{(1)}\hat\varphi^{(0)}\hat\rho^{(0)}_1+v_1^{(0)}\hat\varphi^{(0)}\hat\rho^{(1)}_1\right)+\partial_t\left(\hat\varphi^{(0)}\hat\rho^{(0)}_1\right)\nonumber\\
&\qquad-\partial_z\left(k_1^+\hat\varphi^{(0)}\hat\rho^{(1)}_1-z\hat\varphi^{(0)}\left(\partial_t\hat\rho^{(0)}_1-F_1+\left(k_1^+-D_1\kappa\right)\partial_z\hat\rho^{(1)}_1-D_1\partial_{ss}\hat\rho^{(0)}_1\right)\right)
\nonumber\\
&=D_1\left(\kappa\hat\varphi^{(0)}\partial_z\hat\rho^{(1)}_1+\hat\varphi^{(0)}\partial_{ss}\hat\rho^{(0)}_1\right)
+k_1^+\left(\hat\rho^{(1)}_1-\hat g^{(1)} \right) \partial_z\hat\varphi^{(0)}+\hat\varphi^{(0)}F_1,
\nonumber\\
&\qquad-\partial_z\left(k_1^+\hat\varphi^{(0)}\hat\rho^{(1)}_1-z\hat\varphi^{(0)}\left(\partial_t\hat\rho^{(0)}_1-F_1+\left(k_1^+-D_1\kappa\right)\partial_z\hat\rho^{(1)}_1-D_1\partial_{ss}\hat\rho^{(0)}_1\right)\right).
\nonumber\\
\label{inner 3aa}
\end{eqnarray}
where we have used that $\partial_t\rho^{(0)}_1=\partial_t\hat\rho^{(0)}_1-v^{(0)}\partial_z\hat\rho^{(1)}_1$. Next, after a series of calculations, we rewrite Eq. (\ref{inner 3aa}) as
\begin{eqnarray}
-&\partial_z\left(D_1\hat\varphi^{(0)}\partial_z\hat\rho^{(2)}_1+v_1^{(1)}\hat\varphi^{(0)}\hat\rho^{(0)}_1+v_1^{(0)}\hat\varphi^{(0)}\hat\rho^{(1)}_1+k_1^+\hat\varphi^{(0)}\hat\rho^{(1)}_1\right.
\nonumber\\
&\qquad-z\hat\varphi^{(0)}\left.\left(\partial_t\hat\rho^{(0)}_1-F_1+\left(k_1^+-D_1\kappa\right)\partial_z\hat\rho^{(1)}_1-D_1\partial_{ss}\hat\rho^{(0)}_1\right)\right)
\nonumber\\
&=-k_1^+\hat g^{(1)}\partial_z\hat\varphi^{(0)}
\nonumber\\
&\qquad\qquad +z\partial_z\hat\varphi^{(0)}\left(\partial_t\hat\rho^{(0)}_1-F_1+\left(k_1^+-D_1\kappa\right)\partial_z\hat\rho^{(1)}_1-D_1\partial_{ss}\hat\rho^{(0)}_1\right)
\nonumber\\
\label{inner 3aaa}
\end{eqnarray}
where we have also assumed that $\partial_t\hat\varphi^{(0)}=0$, which will be shown later.
Integrating Eq. (\ref{inner 3aaa}) in $z$ from $-\infty$ to $+\infty$, using the matching conditions and that $\partial_z\hat\rho^{(1)}_1$ is independent of $z$ from Eqs. (\ref{inner 2a}) and (\ref{inner 2c}), we obtain
\begin{eqnarray}
-D_1\nabla\rho^{(1)}_1\cdot\mathbf{n}-v^{(0)}\rho^{(1)}_1=v^{(1)}\rho^{(0)}_1+k_1^+\left(\rho^{(1)}_1-g^{(1)}\right).
\label{bc inner 1}
\end{eqnarray}
An analogous argument can be performed to show that
\begin{eqnarray}
D_0\nabla\rho^{(1)}_0\cdot\mathbf{n}+v^{(0)}\rho^{(1)}_0=-v^{(1)}\rho^{(0)}_0+k_1^+\left(\rho^{(1)}_0-g^{(1)}\right).
\label{bc inner 2}
\end{eqnarray}
Assuming that $v^{(1)}=0$ and $g^{(1)}=0$, facts that we will prove later, we can then conclude that $\rho^{(1)}_1=\rho^{(1)}_0=0$ since these are the unique solutions of Eqs. (\ref{diff 1 pert}) and (\ref{bc inner 1})-(\ref{bc inner 2}).

Next, we analyze the Cahn-Hilliard-like system Eq. (\ref{ch 1})-(\ref{ch 4}). At the outer scale, Eqs. (\ref{ch 1})-(\ref{ch 2}) yield $0=0$ to all orders in $\epsilon$ because $\varphi=0$ or $1$ to all orders. The profiles of $\varphi$ across $\Gamma_1$ and the normal velocity are solely determined from inner expansions. At leading order in the inner scale $O(\epsilon^{-4})$, we obtain
\begin{eqnarray}
&\partial_z\left(G(\hat\varphi^{(0)})\partial_z\hat\mu^{(0)}\right)=0,
\label{ch inner 1}\\
&\hat\mu^{(0)}=B'(\hat\varphi^{(0)})-\partial_{zz}\hat\varphi^{(0)}.
\label{ch inner 2}
\end{eqnarray}
From the matching conditions, we conclude that 
\begin{eqnarray}
&\hat\mu^{(0)}=0,
\label{ch inner 22}\\
&\hat\varphi^{(0)}=1/2\left(1-\tanh{3z}\right).
\label{ch inner 23}
\end{eqnarray}
Observe that $\partial_t \hat\varphi^{(0)}=0$ as assumed earlier. At the next order $O(\epsilon^{-3})$, we obtain
\begin{eqnarray}
&\partial_z\left(G(\hat\varphi^{(0)})\partial_z\hat\mu^{(1)}\right)=0,
\label{ch inner 3}\\
&\hat\mu^{(1)}=B''(\hat\varphi^{(0)})\hat\varphi^{(1)}-\partial_{zz}\hat\varphi^{(1)}-\kappa\partial_z\hat\varphi^{(0)}.
\label{ch inner 4}
\end{eqnarray}
From the matching conditions, we conclude that $\partial_z\hat\mu^{(1)}=0$ so that $\hat\mu^{(1)}=\hat\mu^{(1)}(s,t)$. Multiplying Eq. (\ref{ch inner 4}) by  $\hat\varphi^{(0)}$ and integrating from $-\infty$ to $+\infty$ in $z$, we obtain
\begin{eqnarray}
&\hat\mu^{(1)}(s,t)=\kappa(s,t),
\label{ch inner 4a}\\
&\hat\varphi^{(1)}(s,z,t)=\frac{\kappa(s,t)}{36}\left(1-{\rm sech}^2{3z}\right),
\label{ch inner 5}
\end{eqnarray}
where we have used that $\int^{+\infty}_{-\infty}(\partial_z\hat\varphi^{(0)})^2~dz=1$.
At the next order, $O(\epsilon^{-2})$, we obtain
\begin{eqnarray}
&\partial_z\left(G(\hat\varphi^{(0)})\partial_z\hat\mu^{(2)}\right)=0,
\label{ch inner 6}\\
&\hat\mu^{(2)}=B''(\hat\varphi^{(0)})\hat\varphi^{(2)}+\frac{1}{2}B'''(\hat\varphi^{(0)})(\hat\varphi^{(1)})^2-\partial_{zz}\hat\varphi^{(2)}-\kappa\partial_z\hat\varphi^{(1)}+z\kappa^2\partial_z\hat\varphi^{(0)}.
\nonumber\\
\label{ch inner 7}
\end{eqnarray}
From the matching conditions, we also conclude that $\partial_z\hat\mu^{(2)}=0$ and $\hat\mu^{(2)}=\hat\mu^{(2)}(s,t)$. Multiplying Eq. (\ref{ch inner 7}) by $\partial_z\hat\varphi^{(0)}$ and integrating from $-\infty$ to $+\infty$ in $z$, we obtain
\begin{eqnarray}
\hat\mu^{(2)}(s,t)=\int_{-\infty}^{+\infty}B''(\hat\varphi^{(0)})\hat\varphi^{(1)} \partial_z \hat\varphi^{(1)} ~dz,
\label{ch inner 7a}
\end{eqnarray} 
where we have integrated by parts and used that
\begin{eqnarray}
&0=\int_{-\infty}^{+\infty}\partial_z\hat\varphi^{(0)} \partial_z \hat\varphi^{(1)}~dz,
\label{int 1}\\
&0=\int_{-\infty}^{+\infty}z (\partial_z \hat\varphi^{(0)})^2~dz,
\label{int 2}\\
&0=\int_{-\infty}^{+\infty} \left(B''(\hat\varphi^{(0)})-\partial_{zz} \hat\varphi^{(2)}\right)\partial_z\hat\varphi^{(0)}~dz.
\label{int 3}
\end{eqnarray}
Next, from Eqs. (\ref{ch inner 4}) and (\ref{ch inner 4a}) observe that 
\begin{eqnarray}
B''(\hat\varphi^{(0)})\hat\varphi^{(1)}\partial_z\hat\varphi^{(1)}=\frac{1}{2}\partial_z\left(\partial_z\hat\varphi^{(1)}\right)^2+\kappa\partial_z\hat\varphi^{(1)}\partial_z\hat\varphi^{(0)}+\kappa\partial_z\hat\varphi^{(1)}.
\label{ch inner 7b}
\end{eqnarray}
Combining Eqs. (\ref{ch inner 7a}) and (\ref{ch inner 7b}), we conclude that
\begin{equation}
\hat\mu^{(2)}(s,t)=0
\label{ch inner 7c}
\end{equation}
since $\int_{-\infty}^{+\infty}\partial_z\hat\varphi^{(1)}~dz=0$.

At the next order $O(\epsilon^{-1})$, we obtain
\begin{eqnarray}
&-v_1^{(0)}\partial_z\hat\varphi^{(0)}=\beta\partial_z\left(G(\hat\varphi^{(0)})\partial_z\hat\mu^{(3)}\right)+\beta G(\hat\varphi^{(0)})\partial_{ss}\hat\mu^{(1)}
\nonumber\\
&\qquad\qquad\qquad-\partial_z\hat\varphi^{(0)}\left(k_1^+\left(\hat\rho_1^{(0)}-\hat g^{(0)}\right)+k_1^-\left(\hat\rho_0^{(0)}-\hat g^{(0)}\right)\right)
\label{ch inner 8}
\end{eqnarray}
Integrating Eq. (\ref{ch inner 8}) from $-\infty$ to $+\infty$ in $z$, we obtain
\begin{eqnarray}
v_1^{(0)}&=\beta\partial_{ss}\kappa+k_1^+\left(\rho_1^{(0)}-\hat g^{(0)}\right)+k_1^-\left(\rho_0^{(0)}-\hat g^{(0)}\right)
\label{ch inner 9}
\end{eqnarray}
where we have used that $\hat\mu^{(1)}=\kappa$ from Eq. (\ref{ch inner 4a}).
Next, from Eqs. (\ref{d diff 3}) and (\ref{ch inner 4a}) we obtain
\begin{equation}
\hat g^{(0)}=\rho^*\left(-\mathcal{E}+\tilde\gamma_1\kappa\right).
 \label{ch inner 10}
 \end{equation}
Using these in Eq. (\ref{ch inner 9}), we obtain
\begin{eqnarray}
v_1^{(0)}&=\beta\partial_{ss}\kappa+k_1^+\left(\rho_1^{(0)}- \rho^*\left(-\mathcal{E}+\tilde\gamma_1\kappa\right)\right)
+k_1^-\left(\rho_0^{(0)}-\rho^*\left(-\mathcal{E}+\tilde\gamma_1\kappa\right)\right),
\nonumber\\
\label{ch inner 10a}
\end{eqnarray}
which recovers the sharp interface velocity in Eq. (\ref{s normal vel}). Thus, at leading order we recover the original sharp interface system. Finally, we move to the next order $O(1)$. Here, we obtain
\begin{eqnarray}
&-v_1^{(1)}\partial_z\hat\varphi^{(0)}-v_1^{(0)}\partial_z\hat\varphi^{(1)}=\beta\partial_z\left(G(\hat\varphi^{(0)})\partial_z\hat\mu^{(4)}\right)+\beta\partial_z\left(G'(\hat\varphi^{(0)})\hat\varphi^{(1)}\partial_z\hat\mu^{(3)}\right)
\nonumber\\
&\qquad\qquad\qquad\qquad\qquad\quad+\beta\kappa G(\hat\varphi^{(0)})\partial_z\hat\mu^{(3)}+\beta\partial_s\left(G'(\hat\varphi^{(0)})\hat\varphi^{(1)}\partial_s\hat\mu^{(1)}\right)
\nonumber\\
&\qquad\qquad\qquad\qquad\qquad\quad-\beta zG(\hat\varphi^{(0)})\left(\partial_{ss}\hat\mu^{(1)}+3\partial_s\kappa\partial_s\hat\mu^{(1)}\right)
\nonumber\\
&\qquad\qquad\qquad\qquad\qquad\quad-\partial_z\hat\varphi^{(0)}\left(k_1^+\left(\hat\rho_1^{(1)}-\hat g^{(1)}\right)+k_1^-\left(\hat\rho_0^{(1)}-\hat g^{(1)}\right)\right)
\nonumber\\
&\qquad\qquad\qquad\qquad\qquad\quad-\partial_z\hat\varphi^{(1)}\left(k_1^+\left(\hat\rho_1^{(0)}-\hat g^{(0)}\right)+k_1^-\left(\hat\rho_0^{(0)}-\hat g^{(0)}\right)\right)
\nonumber\\
\label{ch inner 11}
\end{eqnarray}
Integrating Eq. (\ref{ch inner 11}) in $z$ from $-\infty$ to $+\infty$, we obtain
\begin{eqnarray}
-\int_{-\infty}^{+\infty}\partial_z\hat\varphi^{(0)}\left(v_1^{(1)}-k_1^+\hat\rho_1^{(1)}-k_1^-\hat\rho_0^{(1)}\right)~dz
=\beta\kappa\int_{-\infty}^{+\infty} G(\hat\varphi^{(0)})\partial_z\hat\mu^{(3)}~dz,
\nonumber\\
\label{ch inner 12}
\end{eqnarray}
where we have used that $G(0)=G(1)=G'(0)=G'(1)=0$ and
\begin{eqnarray}
&\hat g^{(1)}=\rho^*\tilde\gamma_1\hat\mu^{(2)}=0
\label{ch inner 12aa}\\
&0=\int_{-\infty}^{+\infty}\partial_z\hat\varphi^{(1)}~dz,
\label{ch inner 12a}\\
&0=\int_{-\infty}^{+\infty}G'(\hat\varphi^{(0)})\hat\varphi^{(1)}~dz=\int_{-\infty}^{+\infty}G'(\hat\varphi^{(0)})\partial_s\hat\varphi^{(1)}~dz,
\nonumber\\
\label{ch inner 12b}
&0=\int_{-\infty}^{+\infty}zG(\hat\varphi^{(0)})~dz.
\label{ch inner 12c}
\end{eqnarray}
To make further progress, we observe that
\begin{eqnarray}
&G(\hat\varphi^{(0)})=\partial_z M(\hat\varphi^{(0)}),~~~\rm{where}
\label{ch inner 12d}\\
& M(\hat\varphi^{(0)})=2\left(\hat\varphi^{(0)}\right)^3-3\left(\hat\varphi^{(0)}\right)^2.
\label{ch inner 12e}
\end{eqnarray}
Using these in Eq. (\ref{ch inner 8}), together with the matching conditions, we obtain:
\begin{equation}
\beta G(\hat\varphi^{(0)})\partial_z\hat\mu^{(3)}=\beta\left(M(\hat\varphi^{(0)})+\hat\varphi^{(0)}\right)\partial_{ss}\kappa.
\label{ch inner 13}
\end{equation}
A direct calculation shows that
\begin{equation}
\int_{-\infty}^{+\infty} \left(M(\hat\varphi^{(0)})+\hat\varphi^{(0)}\right)~dz=0.
\label{ch inner 14}
\end{equation}
Combining this with Eq. (\ref{ch inner 12}) we obtain
\begin{equation}
0=\int_{-\infty}^{+\infty}\partial_z\hat\varphi^{(0)}\left(v_1^{(1)}-k_1^+\hat\rho_1^{(1)}-k_1^-\hat\rho_0^{(1)}\right)~dz
\label{ch inner 15}
\end{equation}
Next, from Eqs. (\ref{inner 3c}) and (\ref{inner 3d}), and the matching conditions, we have
\begin{eqnarray}
&\hat\rho^{(1)}_1=\rho^{(1)}_1(s,t)-\frac{1}{D_1}z\left(v_1^{(0)}\hat\rho^{(0)}_1+k_1^+\left(\hat\rho^{(0)}_1-\hat g^{(0)}\right)\right),
\label{ch inner 16}\\
&\hat\rho^{(1)}_0=\rho^{(1)}_0(s,t)-\frac{1}{D_0}z\left(v_1^{(0)}\hat\rho^{(0)}_0+k_1^-\left(\hat\rho^{(0)}_0-\hat g^{(0)}\right)\right).
\label{ch inner 17}
\end{eqnarray}
Using Eqs. (\ref{ch inner 16}) and (\ref{ch inner 17}) in Eq. (\ref{ch inner 15}), we conclude that
\begin{equation}
v_1^{(1)}=k_1^+\rho^{(1)}_1(s,t)+k_1^-\rho^{(1)}_0(s,t).
\label{ch inner 18}
\end{equation}
Finally, using Eq. (\ref{ch inner 18}) in Eqs. (\ref{bc inner 1}) and (\ref{bc inner 2}), we obtain
\begin{eqnarray}
&-D_1\nabla\rho^{(1)}_1\cdot\mathbf{n}-v^{(0)}\rho^{(1)}_1=k_1^+\rho^{(1)}_1\left(1+\rho^{(0)}_1\right)+k_1^-\rho^{(1)}_0\rho^{(0)}_1,
\label{bc inner 1a}\\
&D_0\nabla\rho^{(1)}_0\cdot\mathbf{n}+v^{(0)}\rho^{(1)}_0=k_1^-\rho^{(1)}_0\left(1-\rho^{(0)}_0\right)-k_1^+\rho^{(1)}_1\rho^{(0)}_0.
\label{bc inner 2a}
\end{eqnarray}
We can therefore conclude that $\rho^{(1)}_1=\rho^{(1)}_0=0$, since these are the unique solutions of Eqs. (\ref{diff 1 pert}) and (\ref{bc inner 1a})-(\ref{bc inner 2a}), and that $v_1^{(1)}=0$. Thus, in the region where the outer expansion is valid, we have shown
\begin{eqnarray}
&\rho^\epsilon_1=\rho_1+O(\epsilon^2),
\label{2nd order 1}\\
&\rho^\epsilon_0=\rho_0+O(\epsilon^2),
\label{2nd order 2}\\
&v_1^\epsilon=v_1+O(\epsilon^2),
\label{2nd order 3}
\end{eqnarray}
which demonstrates that the DDM (\ref{d diff 1})-(\ref{ch 2}) provides a 2nd order accurate approximation in $\epsilon$ to the sharp interface model.

\section{Details of the numerical method and implementation}\label{app-methods}
\paragraph{Numerical method}  We use the Crank-Nicolson scheme to discretize the fully time-dependent system Eqs.~(\ref{p1222})-(\ref{far field bc}) in time on larger square domain $\tilde{\Omega}$. In particular, we let $\delta t > 0$ denote the time step, and assume that $\rho_{0}^{n}$, $\rho_{1}^{n}$, $\rho_{0}^{n}$ and $\mu^{n}$ are the solutions at time $t = n\delta t$. We then find the solutions at time $t = (n+1)\delta t$: $\rho_{0}^{n+1}$, $\rho_{1}^{n+1}$, $\rho_{0}^{n+1}$ and $\mu^{n+1}$ by solving
\begin{eqnarray}
&\frac{\varphi_{\infty} H_{0}(\varphi^{n+1})\rho^{n+1}_{0}-\varphi_{\infty} H_{0}(\varphi^{n})\rho^{n}_{0}}{\delta t}=\frac{1}{2}\bigg\{{\bf{\nabla}}\cdot\big (\varphi_{\infty} H_{0}(\varphi^{n+1})D_0(\varphi^{n+1}){\bf{\nabla}}\rho^{n+1}_{0}\big)\nonumber\\
&+\varphi_{\infty}H_{0}(\varphi^{n+1})F_{0}(\varphi^{n+1})-\varphi_{\infty}H_{0}(\varphi^{n+1})\tau_{d}^{-1} \rho^{n+1}_{0}\nonumber\\
&-\varphi_{\infty}|{\bf{\nabla}}\varphi|^{n+1} k_0(\varphi^{n+1})\big[\rho^{n+1}_{0}-\rho^{*}(\mathcal{E}(\varphi^{n+1})+\epsilon^{-1}\gamma(\varphi^{n+1}) \mu^{n+1})\big]\bigg\}\nonumber\\
&+\frac{1}{2}\bigg\{{\bf{\nabla}}\cdot\big (\varphi_{\infty} H_{0}(\varphi^{n})D_0(\varphi^{n}){\bf{\nabla}}\rho^{n}_{0}\big)+\varphi_{\infty}H_{0}(\varphi^{n})F_{0}(\varphi^{n})-\varphi_{\infty}H_{0}(\varphi^{n})\tau_{d}^{-1} \rho^{n}_{0}\nonumber\\
&-\varphi_{\infty}|{\bf{\nabla}}\varphi|^{n} k_0(\varphi)\big[\rho^{n}_{0}-\rho^{*}(\mathcal{E}(\varphi^{n})+\epsilon^{-1}\gamma(\varphi^{n}) \mu^{n})\big]\bigg\},\label{double-dis-q1}\\
&\frac{\varphi_{\infty}H_{1}(\varphi^{n+1})\rho^{n+1}_{1}-\varphi_{\infty}H_{1}(\varphi^{n})\rho^{n}_{1}}{\delta t}=\frac{1}{2}\bigg\{{\bf{\nabla}}\cdot\big (\varphi_{\infty} H_{1}(\varphi^{n+1})D_{1} {\bf{\nabla}}\rho^{n+1}_{1}\big)\nonumber\\
&+\varphi_{\infty}H_{1}(\varphi^{n+1}) F_{1} - \varphi_{\infty}H_{1}(\varphi^{n+1}) \tau_{d}^{-1} \rho^{n+1}_{1}-\varphi_{\infty}|{\bf{\nabla}}\varphi|^{n+1} k_1(\varphi^{n+1})\big[\rho^{n+1}_{1}\nonumber\\
&-\rho^{*}(\mathcal{E}(\varphi^{n+1})+\epsilon^{-1}\gamma(\varphi^{n+1}) \mu^{n+1})\big]\bigg\}\nonumber\\
&+\frac{1}{2}\bigg\{{\bf{\nabla}}\cdot\big (\varphi_{\infty} H_{1}(\varphi^{n})D_{1} {\bf{\nabla}}\rho^{n}_{1}\big)+\varphi_{\infty}H_{1}(\varphi^{n}) F_{1}- \varphi_{\infty}H_{1}(\varphi^{n}) \tau_{d}^{-1} \rho^{n}_{1}\nonumber\\
&-\varphi_{\infty}|{\bf{\nabla}}\varphi|^{n} k_1(\varphi^{n})\big[\rho^{n}_{1}-\rho^{*}(\mathcal{E}(\varphi^{n})+\epsilon^{-1}\gamma(\varphi^{n}) \mu^{n})\big]\bigg\},\label{double-dis-q2}\\
&\frac{\varphi^{n+1}-\varphi^{n}}{\delta t}=\frac{1}{2}\bigg\{\epsilon^{-2}\beta{\bf{\nabla}} \cdot (B(\varphi^{n+1}){\bf{\nabla}}\mu^{n+1})+\epsilon^{-1}\beta{\bf{\nabla}} \cdot (B(\varphi^{n}){\bf{\nabla}}\mu^{n})\bigg\}\nonumber\\
&+\frac{1}{2}\bigg\{|{\bf{\nabla}}\varphi|^{n+1}\big[k_{0}(\varphi^{n+1})\big(\rho^{n+1}_{0}-\rho^{*}(\mathcal{E}(\varphi^{n+1})+\epsilon^{-1}\gamma(\varphi^{n+1})\mu^{n+1})\big)\nonumber\\
&+k_{1}(\varphi^{n+1})\big(\rho^{n+1}_{1}-\rho^{*}(\mathcal{E}(\varphi^{n+1})+\epsilon^{-1}\gamma(\varphi^{n+1})\mu^{n+1})\big)\big]\bigg\}\nonumber\\
&+\frac{1}{2}\bigg\{|{\bf{\nabla}}\varphi|^{n}\big[k_{0}(\varphi^{n})\big(\rho^{n}_{0}-\rho^{*}(\mathcal{E}(\varphi^{n})+\epsilon^{-1}\gamma(\varphi^{n})\mu^{n})\big)\nonumber\\
&+k_{1}(\varphi^{n})\big(\rho^{n}_{1}-\rho^{*}(\mathcal{E}(\varphi^{n})+\epsilon^{-1}\gamma(\varphi^{n})\mu^{n})\big)\big]\bigg\},\label{double-dis-p1}\\
&\mu^{n+1}=-\epsilon^{2} \Delta\varphi^{n+1}+G'(\varphi^{n+1}),\label{double-dis-mu}
\end{eqnarray}	
with the following boundary conditions
\begin{eqnarray}
{\bf{\nabla}}\rho^{n+1}_{0}\cdot {\bf{n}}={\bf{\nabla}}\rho^{n+1}_{1}\cdot {\bf{n}}={\bf{\nabla}}\varphi^{n+1}\cdot {\bf{n}}={\bf{\nabla}}\mu^{n+1}\cdot {\bf{n}}=0~~~~{\rm on}~~\partial \tilde{\Omega}.
\end{eqnarray}
Moreover, we add a small positive parameter $\delta = 10^{-5}$ to the functions $H_0$, $H_1$ and $B(\varphi)$ in all second-order differential operators in (\ref{double-dis-q1})-(\ref{double-dis-mu}) as a regularization.
\paragraph{Implementation} Standard, cell-centered central-difference finite difference methods are used, together with a block-structured adaptive mesh, to discretize the equations in space.
The nonlinear equations at the implicit time level are solved using an efficient nonlinear FAS multigrid solver. See  \cite{BSAM20} for details. Here, we use a 4-level block-structured adaptive mesh, which consists of one root level (grid size $h_{0}$) and three refinement levels (grid size $h_{i}$) with refinement ratio of 2. For each adaptive mesh level, we refine the grid cell ($i,j$) wherever $h_{i}|{\bf{\nabla}}\varphi_{i,j}|>q_{tol}$. Here, we set $q_{tol}=0.01$.

\section{Convergence of anisotropic layer dynamics}
\label{convergence_appendix}

Here we present the convergence analysis using the fully time-dependent dynamics. The results for quasi-steady dynamics are similar (not shown). Using the parameters in Sec. \ref{anisotropic dynamics}, we analyze the convergence of our schemes at time $t=0.1$. The consecutive errors (e.g., Eq. (\ref{consecutive errors})) and convergence rates for the adatom concentrations are given in Tables \ref{tab111-conv-6-1} and \ref{tab112-conv-3-2} for 6-fold and 3-fold symmetric anisotropic edge energies and kinetic coefficients, respectively. The results suggest the scheme is second-order convergent in both the ${\ell}_{2}$ and ${\ell}_{\infty}$ norms.

\begin{table}[h]
\caption{Convergence test for adatom concentrations $\rho_{2}$, $\rho_{1}$ and $\rho_{0}$ using 6-fold symmetric anisotropic edge energies and kinetic coefficients in \S \ref{anisotropic dynamics}.}\label{tab111-conv-6-1}
\begin{indented}
\lineup
\item[]
	\begin{tabular}{c||c|c||c|c||c|c}
	\br
 t=0.1&${\ell}_{2}$&&&&&\\
	\hline
$\epsilon$     & $E^{(2)}_{\epsilon_{i-1},\epsilon_i,\rho_{2}}$ & rate& $E^{(2)}_{\epsilon_{i-1},\epsilon_i,\rho_{1}}$ & rate& $E^{(2)}_{\epsilon_{i-1},\epsilon_i,\rho_{0}}$ & rate
	\\
	\hline
0.4 &1.172$\times 10^{-3}$    & ---&  1.642$\times 10^{-3}$& ---&  1.007$\times 10^{-3}$ & ---
	\\
	\hline
0.2 &1.143$\times 10^{-3}$ & 0.04& 1.454$\times 10^{-3}$ & 0.18& 1.065$\times 10^{-3}$  & -0.08
	\\
	\hline
0.1 &6.485$\times 10^{-4}$ & 0.82& 8.359$\times 10^{-4}$ & 0.80& 7.196$\times 10^{-4}$ & 0.57
	\\
	\hline
0.05 &1.942$\times 10^{-4}$ &1.74 & 2.631$\times 10^{-4}$ & 1.67&2.734$\times 10^{-4}$ &1.40 
	\\
	\hline
0.025 &5.304$\times 10^{-5}$ &1.87 & 6.413$\times 10^{-5}$ & 2.04&7.654$\times 10^{-5}$ &1.84 \\
\hline
 t=0.1&${\ell}_{\infty}$&&&&&\\
	\hline
$\epsilon$     & $E^{(\infty)}_{\epsilon_{i-1},\epsilon_i,\rho_{2}}$ & rate& $E^{(\infty)}_{\epsilon_{i-1},\epsilon_i,\rho_{1}}$ & rate& $E^{(\infty)}_{\epsilon_{i-1},\epsilon_i,\rho_{0}}$ & rate
	\\
	\hline
0.4 &2.255$\times 10^{-3}$ & ---& 3.240$\times 10^{-3}$& ---&  4.006$\times 10^{-3}$ & ---
	\\
	\hline
0.2 & 3.491$\times 10^{-3}$ & -0.63& 4.574$\times 10^{-3}$ & -0.50&  5.693$\times 10^{-3}$& -0.50
	\\
	\hline
0.1 & 2.345$\times 10^{-3}$ & 0.57& 3.233$\times 10^{-3}$ & 0.50& 4.758$\times 10^{-3}$& 0.26
	\\
	\hline
0.05 &7.939$\times 10^{-4}$  &1.56 &1.413$\times 10^{-3}$  &1.19 &1.848$\times 10^{-3}$ &1.36
	\\
	\hline
0.025 &1.823$\times 10^{-3}$  &2.12 &4.094$\times 10^{-4}$  &1.79 &5.537$\times 10^{-4}$ &1.74\\
\br
\end{tabular}
\end{indented}
\end{table}

\begin{table}
\caption{Convergence test for adatom concentrations $\rho_{2}$, $\rho_{1}$ and $\rho_{0}$ using 3-fold symmetric anisotropic edge energies and kinetic coefficients in \S \ref{anisotropic dynamics}.}\label{tab112-conv-3-2}
\begin{indented}
\lineup
\item[]
	\begin{tabular}{c||c|c||c|c||c|c}
\br 
 t=0.1&$\ell_{2}$&&&&&\\
	\hline
$\epsilon$     & $E^{(2)}_{\epsilon_{i-1},\epsilon_i,\rho_{2}}$ & rate& $E^{(2)}_{\epsilon_{i-1},\epsilon_i,\rho_{1}}$ & rate& $E^{(2)}_{\epsilon_{i-1},\epsilon_i,\rho_{0}}$ & rate
	\\
	\hline
0.4 & 1.584$\times 10^{-3}$ & ---& 2.700$\times 10^{-3}$& ---& 1.872$\times 10^{-3}$ & ---
	\\
	\hline
0.2 & 1.752$\times 10^{-3}$ & -0.15&  2.253$\times 10^{-3}$& 0.26& 1.685$\times 10^{-3}$ & 0.15
	\\
	\hline
0.1 & 1.016$\times 10^{-3}$ & 0.79&  1.031$\times 10^{-3}$& 1.13& 7.835$\times 10^{-4}$ & 1.10
	\\
	\hline
0.05 & 3.223$\times 10^{-4}$ & 1.66&  2.789$\times 10^{-4}$& 1.89& 1.872$\times 10^{-4}$ & 2.07\\
	\hline
0.025 & 8.852$\times 10^{-5}$ & 1.86&  7.001$\times 10^{-5}$& 1.99& 4.502$\times 10^{-5}$ & 2.06\\
\hline
 t=0.1&$\ell_{\infty}$&&&&&\\
	\hline
$\epsilon$     & $E^{(\infty)}_{\epsilon_{i-1},\epsilon_i,\rho_{2}}$ & rate& $E^{(\infty)}_{\epsilon_{i-1},\epsilon_i,\rho_{1}}$ & rate& $E^{(\infty)}_{\epsilon_{i-1},\epsilon_i,\rho_{0}}$ & rate
	\\
	\hline
0.4 &  7.321$\times 10^{-3}$& ---& 5.747$\times 10^{-3}$& ---& 4.228$\times 10^{-3}$ & ---
	\\
	\hline
0.2 &1.001$\times 10^{-2}$ & -0.45&  6.091$\times 10^{-3}$& -0.08&  4.982$\times 10^{-3}$ & -0.24
	\\
	\hline
0.1 & 5.620$\times 10^{-3}$ & 0.83&  4.623$\times 10^{-3}$& 0.40&2.686$\times 10^{-3}$ & 0.89
	\\
	\hline
0.05 & 1.934$\times 10^{-3}$ & 1.54&  1.940$\times 10^{-3}$& 1.25& 7.734$\times 10^{-4}$ & 1.80
	\\
	\hline
0.025 & 5.500$\times 10^{-4}$ & 1.82&  5.590$\times 10^{-4}$& 1.80& 1.931$\times 10^{-4}$ & 2.00\\
\br
\end{tabular}
\end{indented}
	\end{table} 	

\clearpage

\section*{References}
\bibliography{Reference.bib}{}
\bibliographystyle{plain}

\newpage

\clearpage

\clearpage

\clearpage

\clearpage

\clearpage

\clearpage

\clearpage

\clearpage

\clearpage

\clearpage

\clearpage

\end{document}